\documentclass[apj]{emulateapj}	
\usepackage{apjfonts}		

\usepackage{natbib}
\usepackage{graphicx}
\bibpunct{(}{)}{;}{a}{}{,}

\newcommand{\lina}{\ion{Fe}{1}~$\lambda$6302.5~\AA}
\newcommand{\linb}{\ion{Fe}{1}~$\lambda$6301.5~\AA}
\newcommand{\linc}{\ion{Fe}{1}~$\lambda$15648~\AA}

\newcommand{\paperii}{Paper~{\sc ii}}
\newcommand{\paperi}{Paper~{\sc i}}
\newcommand{\los}{line-of-sight}
\slugcomment{@/scratch/asp/95.10.15/text/ms.tex}
\shorttitle{MISMA inversion of a full Sunspot}
\shortauthors{S\'anchez Almeida}

\begin{document}
   \title{Physical properties of the solar magnetic photosphere
	under the\\ MISMA hypothesis III: sunspot at disk center}

    \author{J.~S\'anchez~Almeida}
    \affil{Instituto de Astrof\'\i sica de Canarias, 
              E-38205 La Laguna, Tenerife, Spain}
   \email{jos@iac.es}

\begin{abstract}
Small-scale fluctuations 
of magnetic field and velocity may be responsible for the 
Stokes asymmetries observed in all photospheric magnetic
structures (the MISMA hypothesis).
We support the hypothesis showing that atmospheres with optically-thin
micro-structure
reproduce the polarization of \linb\ and 
\lina\  observed in a sunspot.
Ten thousand spectra were fitted by model MISMAs with two 
magnetic components 
interleaved along the \los . 
Combining all the fits, we set up a semi-empirical
model sunspot characterized by  two components
with very different magnetic field  inclination.
The major component, which contains most of the mass, is
more vertical than the minor component.
The field lines
of the minor component are inclined below the horizontal
plane throughout the penumbra.
Magnetic field lines and mass
flows are parallel, consequently, both upflows and
downflows are present everywhere on the penumbra.
Major and minor components have very different velocities
(several hundred m~s$^{-1}$ for the major
component versus
10~km~s$^{-1}$ for the minor component), but
the mass transported per unit time is similar.
The similarity between the vertical mass flow and 
the magnetic flux of the two 
components suggests
that field lines  emerging as major component 
may return to the photosphere as minor component.
If so, the observed magnetic field strength difference between
components leads to 
a siphon flow whose magnitude and direction
agree with the inferred Evershed flow.
Several 
tests support the internal consistency 
of the retrieved model sunspot. 
The magnetic field  vector
{\bf B} does not violate the $\nabla {\bf B}=0$ condition.
The model sunspot reproduces the net circular polarization
of the observed lines, plus
the abnormal behavior 
of \linc .
The use of only one magnetic component
to interpret the spectra leads to inferring 
upflows in the inner penumbra and
downflows in the outer penumbra, in agreement with
previous findings. 
\end{abstract}


%
%

\section{Introduction}\label{introduction}

Ubiquitous small-scale fluctuations of the magnetic field and the
velocity may be responsible for the  Stokes
asymmetries\footnote{We will use the standard notation for
the Stokes parameters; $I$ for the intensity,
$Q$ and $U$ for the two independent types
of linear polarization, and $V$ for the
circular polarization. The Stokes profiles are 
the function Stokes parameter versus 
wavelength for a particular spectral line.
These Stokes profiles are expected to be 
symmetric or anti-symmetric when the 
line is formed in a homogeneous 
atmosphere, e.g., \citet{unn56,lan92}. 
Deviations from these
symmetries are called Stokes asymmetries.}
observed in all photospheric magnetic structures. 
If the spatial scale 
is small enough to be
optically-thin,  one has a very efficient source
of line asymmetry.
The existence of such  gradients 
was conjectured
by \citet{san96}, that coined the acronym MISMA
to describe magnetic atmospheres having optically-thin
irregularities (MISMA~$\equiv$~MIcro-Structured Magnetic 
Atmosphere).
If the conjecture turns out to be correct, we must
understand the magnetic micro-structure
for two fundamental reasons.
First, many physical  processes determining the 
magnetism at all scales take place at micro-scales 
(e.g., ohmic diffusion, advection, onset of 
instabilities, viscous stresses, etc.). Second, 
it plays a key role for the proper 
interpretation of the magnetic
field measurements, which are prone to bias if 
the  micro-structure is overlooked.  
(For a more elaborated discussion on 
these two reasons, see \citealt{san01b}.)

This paper is the third of a series aimed at
examining the conjecture in detail. We wanted
to see (a) whether the observed asymmetries can be 
quantitatively explained by
MISMAs, and (b) what are the physical 
properties of the micro-structure.
The first paper introduces
an inversion code (IC)
based on the MISMA concept. It is a (fairly) automatic
algorithm to fit Stokes profiles
using synthetic profiles formed in
atmospheres with optically-thin micro-structure
\citep[][ \paperi]{san97b}.
As a test case, \paperi\ 
reproduces the asymmetries found in plage regions 
by \citet{ste84}.
The second paper describes the successful application 
of the IC to quiet Sun network and inter-network regions
\citep[][ \paperii; see also \citealt{soc02}]{san00}. 
The present paper extends
the MISMA inversions to a sunspot, showing that 
the micro-structure is also able
to cope with the Stokes asymmetries
of sunspots. 

The paper is organized as follows.
The properties of the spectro-polarimetric
data subject to analysis
are summarized in \S~\ref{observations}.
Section~\ref{sec_inv} describes the 
constraints used to set up
the model MISMAs, the quality of the fits,
and the procedure
to assemble independent inversions 
from different pixels into a single coherent
structure.
The main results are set forth in the seven
subsections of \S~\ref{main}.
Specific aspects of the resulting 
model sunspot are examined in \S~\ref{mfc_sect} (magnetic
flux conservation) and \S~\ref{mass} (mass conservation).
Section \S~\ref{role} explains
the physical origin of the asymmetries from a 
radiative transfer point of view.
Finally, \S~\ref{conclusions} comments on and
summarizes the contents 
of the paper. 
The interpretation of these results in the 
context of the global penumbral magnetic field 
topology is  deferred for a 
separate  paper \citep{san04c}.
We have preferred to 
disconnect the fact that
MISMAs account for the Stokes asymmetries
in sunspots (this work), from the
more speculative physical
interpretation of the retrieved 
micro-structure. 
Even if the physical interpretation 
is eventually proved to be wrong, the
micro-structure would still account
for the asymmetries. 

%

\section{Observations and data}\label{observations}

\begin{figure}
\epsscale{1.0}
\plotone{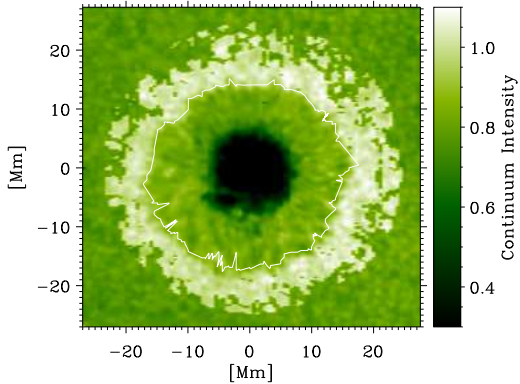}
\caption{Continuum image of the inverted sunspot.
Coordinates are given in Mm from the adopted
sunspot center. 
Those pixels which have been  inverted are
artificially enhanced so that the sunspot and
its surroundings 
stand out against a dark background 
of pixels without inversion.
The white contour shows the border
of the sunspot, and it is used for reference along the paper.
}
\label{cont_image}
\end{figure}

The Stokes $I$, $Q$, $U$ and $V$ 
profiles
analyzed in this work were obtained by \citet{lit98}  with the  
Advanced Stokes Polarimeter \citep[ASP;][]{elm92}.
They correspond to the sunspot in active region NOAA~7912
(Fig.~\ref{cont_image})
observed on 1995 October 15 when passing close to the solar disk center
(S5, E9, cosine of heliocentric angle $\mu=0.96$). The region was
scanned with a step-size of 0\farcs 75 
in the East-West direction, perpendicular
to the slit.
The pixel size along the spectrograph slit
0\farcs~37, is enough to sample a seeing
close to (but better than) 1\arcsec .  
The Field-Of-View (FOV) is set by the length
of the slit, and the size of the scan, and it
turns out to be 85\arcsec~$\times$135\arcsec .
The ASP provides the polarization  of
\linb\ and \lina\  every 12.8~m\AA .
An integration time per slit position
of 10~sec affords a noise per spectral pixel
better than $10^{-3}$ times the continuum
intensity.
The spectra are corrected for instrumental
polarization to a level below this noise \citep{sku97}.
For further details on the 
observational setup and the target, 
refer to \citet{lit98}.
We use this rather old data set because its
image quality was among the best ASP maps
available when the work began in 1999. 
%

The sunspot has negative
polarity, i.e., the magnetic field lines are
anti-parallel to the vertical direction at the
umbral core.
This fact introduces some minor complications in the 
interpretation of the sunspot magnetic field.
They will be pointed out and
clarified whenever they interfere
with our argumentation.

The zero of the absolute velocity scale
has been set using the Stokes $V$
zero-crossing wavelengths 
in the plage around the
sunspot. We use its mean position 
as null velocity.
The same procedure was used in \paperii\ since
observations have shown that
the zero-crossing wavelengths of  
Stokes $V$ profiles in plage are very close to the
laboratory wavelengths of the lines 
\citep{sol86,mar97}. The systematic error of our
absolute velocities should be
less than 0.25~km~s$^{-1}$.

\section{Inversion}\label{sec_inv}

\subsection{MISMA scenario to represent umbrae and
	penumbrae}\label{scenario}
%
\begin{figure}
\includegraphics[scale=.6]{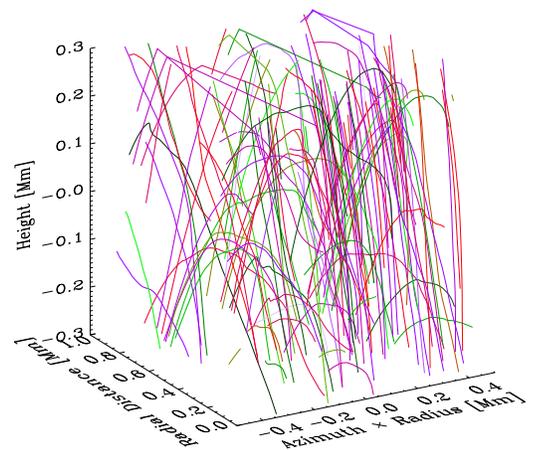}
\includegraphics[scale=.7,clip=true]{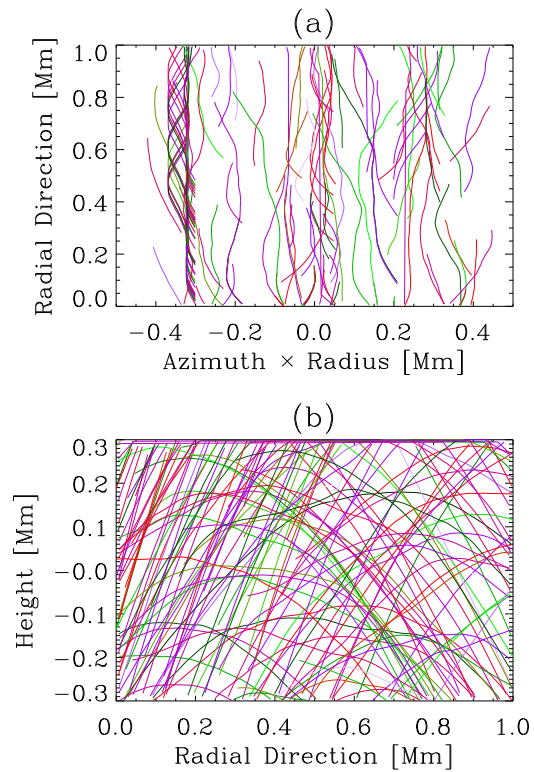}
\caption{Top: cartoon with the kind of atmosphere
used for inversion of sunspot Stokes profiles. 
It represents magnetic field lines in one resolution element.
They overlap along any \los\ . (a) A view from above 
showing how the field lines are preferentially aligned
along the sunspot radial direction. (b) A view from the
side showing field lines with very different vertical
inclinations.}
\label{cartoon}
\end{figure}
%
We employ the scenario proposed in \citet[][ \S~3.1]{san96},
which  resembles a scaled-down version of 
the {\em uncombed fields} of \citet{sol93b}.
The atmosphere has two magnetic components which differ in velocity,
inclination, as well as in many other physical parameters. 
It is a MISMA, therefore, each \los~
crosses several different substructures (see
the schematic in Fig.~\ref{cartoon}). Since we will use 
the Inversion Code~(IC) by \citet{san97b}, these two components meet
several physical requirements.
\begin{enumerate}
\item \label{item2} The temperature of the two magnetic
	components
	is the same at every geometrical height. We adopt
	this approach for consistency,
	since the radiative exchange among the optically-thin
	micro-structures of the MISMA should efficiently smooth
	temperatures in seconds
	\citep[e.g.,][]{spi57,sti91}\footnote{However, 
	this hypothesis makes our model insensitive
	to the temperature variations 
	associated with the smallest  
	penumbral filaments observed by, e.g.,
	\citet{sch02}.}.
\item \label{item1} In addition to the two magnetic 
	components, there is a third non-magnetic component.
	It is introduced for technical reasons and 
	plays no central role. As we will see (\S~\ref{ocf_sect}),
	the fraction of photospheric volume that
	it occupies within the sunspot is always marginal.
\item \label{item3} The three components are in lateral pressure
	balance at every geometrical height (i.e., they have the same total
	pressure, which is the sum of the gas pressure plus the
	magnetic pressure).
	This constraint and the assumption of  hydrostatic
	equilibrium along field lines provide
	the full vertical stratification of magnetic field strengths
	from the strengths at one height.
	The field strengths of the different components are different.
\item \label{item6} The two magnetic components have 
	different inclinations with respect to the 
	vertical direction, but the same azimuth.
	Inclinations and azimuths do not vary with height
	in the atmosphere.
\item \label{item4} Only motions along magnetic field
	lines are allowed. The mass conservation along field lines
	determines the velocity stratification by setting the
	velocity at a given height.
\item \label{item5} Unpolarized stray light is accounted for.
	Dealing with
	stray light introduces three
	free parameters: the fraction of stray light,
	a global velocity, and a microturbulence.
	The thermodynamic stratification is that
	of the non-magnetic component
	of the MISMA model atmosphere. 
\end{enumerate}

\begin{figure*}
\epsscale{.9}
\includegraphics[angle=90,scale=.7]{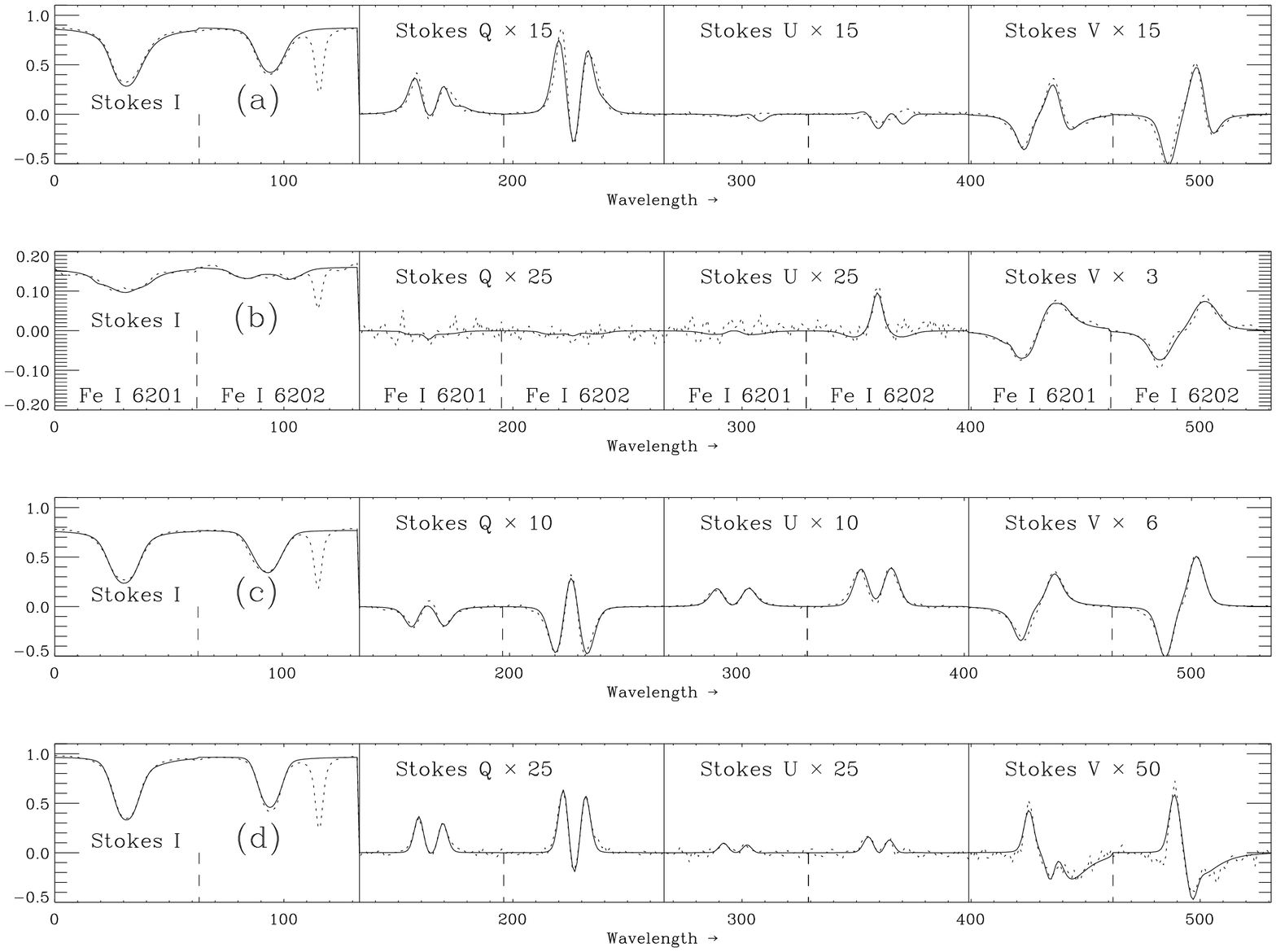}
\caption{Examples of the quality of the fits provided
by the MISMA inversions. They show Stokes $I$, $Q$,
$U$ and $V$ profiles of \linb\ and \lina . (The
different types of profiles are separated by vertical
solid lines, and the division
between \linb\ and \lina\ is indicated by vertical dashed 
lines.) 
The solid lines are synthetic profiles whereas the dotted lines
show observations.
	(a) Profiles in the limb-side penumbra. (b) Profiles
	in the umbra. (c) Profiles in the center-side 
	penumbra. (d) Profiles outside the limb-side penumbra.
	The abscissae represent the pixel number in an array used
	to carry out the fits. Within each type
	of Stokes profile, wavelengths increase to the right,
	with a gap between \linb\ and \lina\ to exclude a telluric
	line.
	Note the large asymmetries of some of the profiles, and the
	fair fit provided by the inversions;  e.g., Stokes $V$
	in (a) and (d).
}
\label{show_fit}
\end{figure*}

The actual model atmospheres are described using 23 parameters: 
the temperature of the non-magnetic component
at four heights,
a  global scaling factor for the temperature of  
the magnetic components
(item \ref{item2} above),
two magnetic field strengths
at the bottom of the atmosphere
(or BID, which stands for Bottom of Integration Domain)
(item~\ref{item3}), 
two occupation fractions at BID
(i.e., the fraction of atmospheric
volume occupied by each magnetic component;
item~\ref{item1}),
two velocities at BID (item~\ref{item4}),
two magnetic field inclinations (item~\ref{item6}),
one magnetic field azimuth (item~\ref{item6}),
four parameters to define the distribution of
microturbulent velocities (two for the magnetic
components,  
and two for the non-magnetic component),
three parameters to account
for the stray light (item \ref{item5}), 
one macro-turbulent velocity (which simulates 
instrumental broadening plus macro-turbulence 
of solar origin) and, finally,
the maximum pressure at BID.
The latter varies to assure that the model atmosphere
embraces the region between $\log\tau_c=1$ and
$\log\tau_c=-4$ (with $\tau_c$ 
the continuum  optical
depth at 6302 \AA). This range of
optical depths guarantees
that the lines 
\lina\ and \linb\ 
are entirely formed within the bounds of the  model atmosphere.
Recovery of the complete atmospheric structure from these 23
parameters is described in
\citet{san97b}\footnote{But note the misprint in equation~(13).}.
These 23 parameters must be set during the least-squares
fitting procedure.
Although the number of free parameters 
seems large, it has to be compared with
the  536 different observed quantities to be fitted 
(67 wavelengths per line 
and per Stokes parameter; see Fig.~\ref{show_fit}).
These observables are not fully independent (e.g., 
the finite spectral resolution links neighbor wavelengths, 
and the radiative transfer couples the four Stokes parameters), but
the author believes that
the number of observables is so large compared with the number of
free parameters that the number of {\em independent} observables
almost certainly exceeds 23.

%

\begin{figure}
\includegraphics[angle=0,scale=.6]{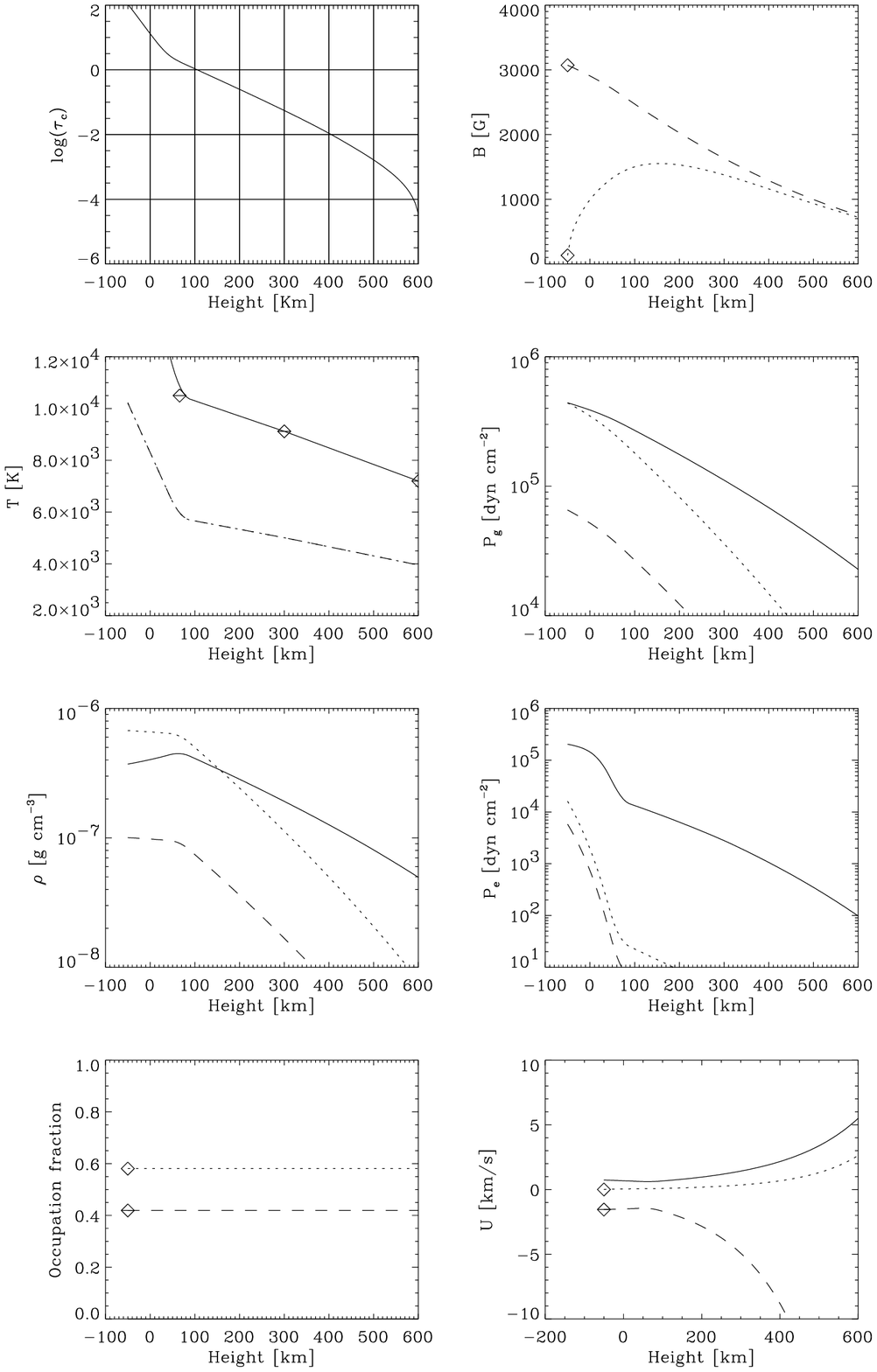}
\caption{Typical model MISMA retrieved
by the IC used in this work. It 
reproduces the profiles in Fig.~\ref{show_fit}a.
The different panels represent the variation
with height in the atmosphere of various physical parameters
(from left to right and from top to bottom):
the
continuum optical depth, the magnetic field strength,
the temperature, the gas pressure, the density, the electron pressure,
the occupation fraction, and the speed along field lines.
Except for the upper-left panel,
the type of line identifies the three
components of the model MISMA; the solid lines correspond
to the non-magnetic component and the dotted
and dashed lines to the magnetic components.
The diamonds indicate the free parameters of the model,
from which the entire stratification of the 
atmosphere has been deduced.
Note that the atmosphere is only meaningful within the range
of optical depths where the observed lines are formed 
(e.g., the drop of major component
magnetic field strength below $\log\tau_c=0$ is
not significant; $\log\tau_c=0$ occurs at 100~km).
}
\label{model_misma}
\end{figure}

%
Figure~\ref{cartoon} illustrates the kind of complex
distribution of magnetic field lines 
that our schematic model MISMAs may be modeling. 
Field lines like these
are represented by our IC as, e.g., the model MISMA 
in Fig.~\ref{model_misma}.
If the micro-structure of the atmosphere
is optically-thin then
the emerging spectrum does not depend on 
details of the atmosphere but only on its 
average properties \citep{san96}. Therefore, describing 
a messy bunch of field lines with 
only two components is consistent with our MISMA
assumption. Obviously, we cannot infer 
details of the micro-structure but volume 
average properties.
In particular, we are insensitive to the size of the
micro-structures, which we consider
to be optically-thin and
so smaller than 100 km.
The cartoon in Fig.~\ref{cartoon} represents magnetic field lines
existing in a single resolution element. 
The field lines are aligned approximately radially
from the sunspot center,
having different azimuths and inclinations.
Figure~\ref{cartoon}a shows a view from the top of the
3-dimensional distribution of field lines.
They show a global radial orientation but, due to slight
misalignments and swings, many different field lines 
overlap along any \los .
The axes,  given in Mm, are for guidance only.
Figure~\ref{cartoon}b represents
a side view of the same distribution of field
lines. Different magnetic field
inclinations coexist in the resolution elements.
%

%
The MISMA scenario and the requirements used for
sunspot inversions are basically those
that successfully
reproduce the Stokes  profiles observed in plage,
network and internetwork regions \citep{san97b,san00,soc02}.
The main difference being the inclination
of the magnetic fields, and the different temperatures
of the magnetic components and the background.
The similarities between scenarios  probably
reflects both the degeneracy of the radiative 
transfer for MISMAs and the fact that  
the employed MISMA  scenario grasps the essential
degrees of freedom required to reproduce
most asymmetries found in the photosphere.
As explained above, there are two magnetic component
in each model MISMA. They will be denoted as 
minor and major depending on which one has largest 
$\alpha\rho$, i.e., the largest product of
occupation fraction $\alpha$ times density $\rho$.
Consequently, the major component is the one 
containing  most of the mass of the atmosphere.
This intuitive classification is also
used in \paperii . Here, however,
the component of largest $\alpha\rho$ 
may change  along the atmosphere, making the definition ambiguous.
Fortunately,
the change occurs in a small fraction of inversions (3\%)
and, even in these cases, it only affects a few superficial
layers. (The classification is based on
$\alpha\rho$ at the bottom of the atmosphere.) 

\subsection{Strategy for inversion}

First, 
the center of the sunspot was set by trial and error.
We minimize the  departures from axi-symmetry of the distribution
of observed continuum intensities.
The chosen center is  1.3 Mm away from the darkest point
in the umbra. The resulting intensity distribution is fairly symmetric,
leading to 
a sunspot radius of about 17~Mm,
which is the value adopted in the paper.

We choose for inversion all those pixels within 
24~Mm  from the sunspot center
having a degree of polarization larger than 0.5\% 
(average polarization over a bandpass  75~m\AA\ wide in 
the blue wing of \lina ).
This criterion yields  10130 different pixels encompassing
the entire sunspot (Fig.~\ref{cont_image}). We have carried
out an inversion for each one of these pixels using
the IC described in \paperi\ and the scenario presented
in \S~\ref{scenario}.
The strategy for inversion is slightly different from the
one in \paperii . We also begin by classifying 
all Stokes profiles using a principal
component analysis scheme.
This  procedure sorts the observed Stokes profiles according
to similarities in shape.
We start the inversion at the sunspot center, and then proceed
with the closest profile according the classification.
However, rather than using the last retrieved 
atmosphere to initialize each new inversion,
we look up in a data base with all previously
inverted synthetic profiles.
The atmosphere of the closest one (in a least squares sense) is chosen
for initialization.
The process is repeated to completion of the 10130 sets
of Stokes profiles.
Each automatic inversion was visually inspected before acceptance.
Failures were re-started from a different
model atmosphere until the agreement was
satisfactory.

\subsection{Quality of the fits}\label{quality}
Figure \ref{show_fit} shows four examples of the quality of the
fits provided by the MISMA inversions. We have chosen
the Stokes profiles at four 
positions on the sunspot, (a) limb-side penumbra,
(b) umbra, (c) center-side penumbra, and (d) outside
the penumbral border defined by the continuum image
(the letters correspond to the labels on the
figure, whereas limb-side and center-side refer 
to the portion of penumbra closest to or farthest from
the solar limb).
These four regions produce Stokes profiles with
the variety of shapes to be reproduced with
the inversions, including severe Stokes asymmetries.
The Stokes $V$ profile in Fig.~\ref{show_fit}a shows
the so-called cross-over effect \citep{gri72,gol74},
with Stokes $V$ having three or more lobes. It
occurs in the neutral line, where the mean magnetic
field is perpendicular to the \los ,  and Stokes $V$ should
be zero if the magnetic field were
spatially resolved \citep{mak86,san92b}.
Despite the  quality of the fits, we would like to
point out that they are not as good as for the
quiet Sun network and internetwork 
profiles described in \paperii\ and \citet{soc02}.
We have not been able to pin down the cause, however,
it obviously implies that the scenario
adopted for inversion suits better the conditions
in the quiet Sun. 

Yet another hint of the  quality of the fits is 
provided by the ability to reproduce the observed 
net circular polarization, a result analyzed
in detail in \S~\ref{ncp_sect}.

\subsection{Removal of the azimuthal ambiguity}\label{ambiguity}

\begin{figure}
\epsscale{1.}
\plotone{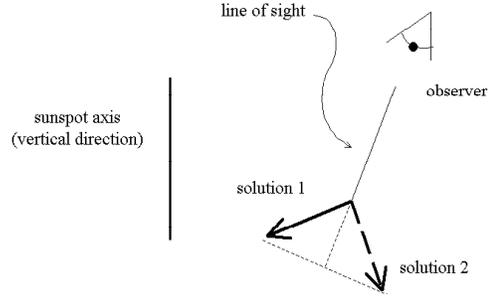}
\caption{Schematic showing the two possible
solutions compatible with a given Zeeman absorption.
The two have the same magnetic field component
along the \los , and differ by 180$^\circ$ in 
azimuth. 
We have tried to select the solution that converges toward
the sunspot axis (solution 1).
(We deal with a negative polarity sunspot, so that
solution 1 tends to be anti-parallel to the vertical 
direction.)
}
\label{schematic2}
\end{figure}

\begin{figure*}
\epsscale{1.}
\plottwo{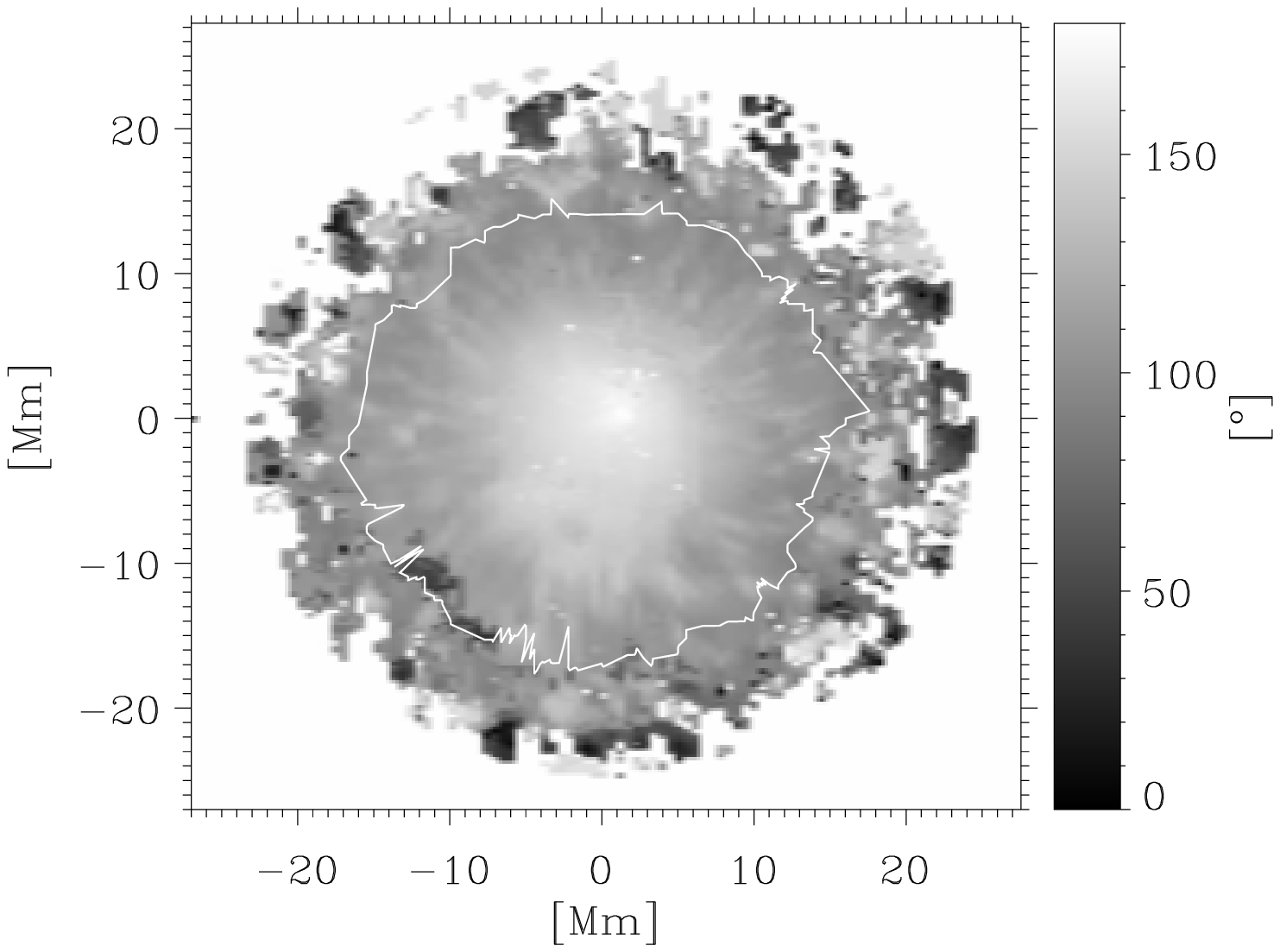}{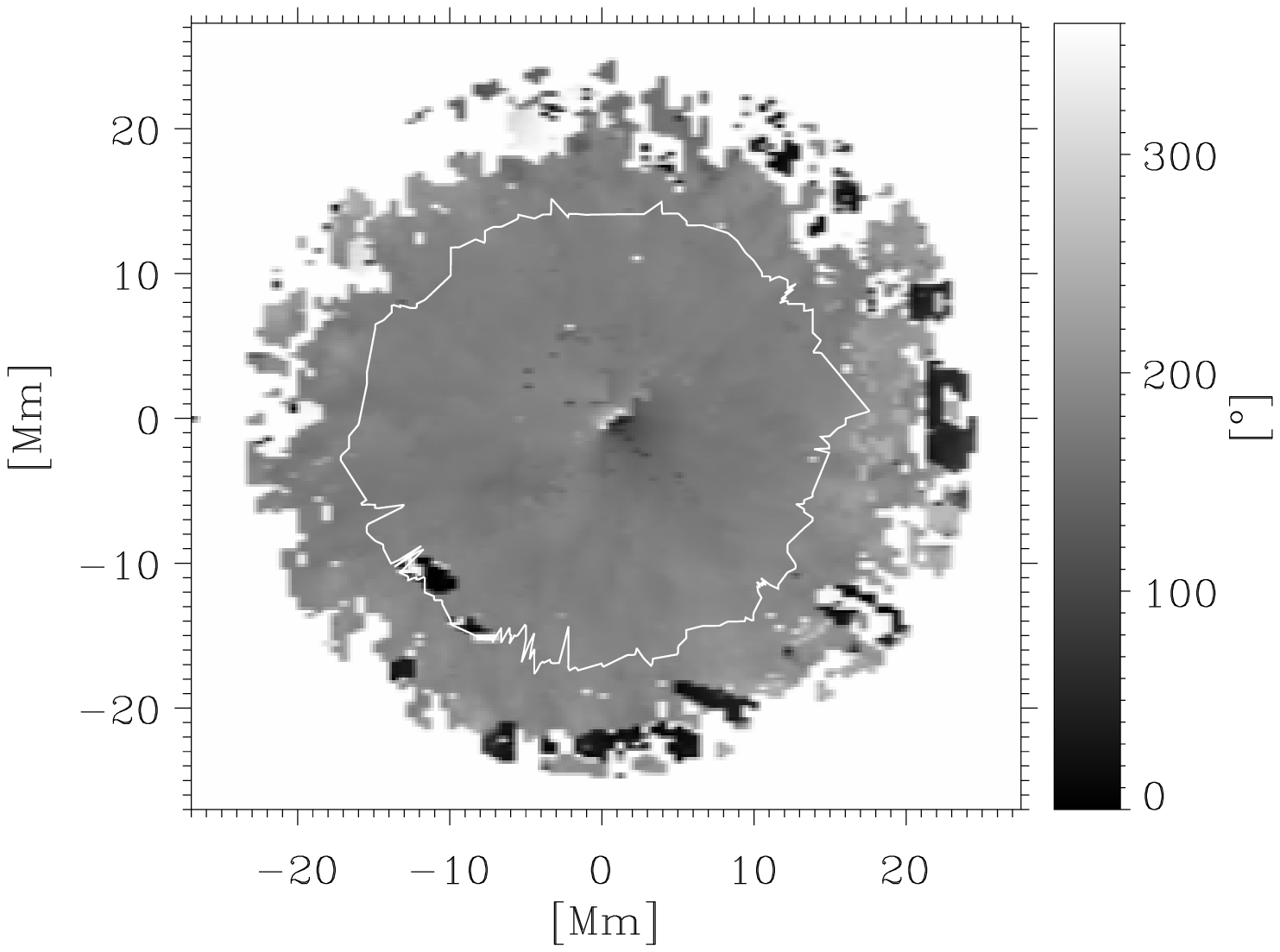}
\caption{Maps of the inclination (left) and the local azimuth (right) of the
major magnetic component of the MISMA inversions.
The inclinations are scaled between 0$^\circ$ and 180$^\circ$
whereas the azimuths  go from
0$^\circ$ to 360$^\circ$.
Note the axi-symmetry of the retrieved magnetic fields.
The white outline 
corresponds to the
external penumbral border (see Fig.~\ref{cont_image}).
}
\label{inc_az_image}
\end{figure*}

\begin{figure*}
\epsscale{1.0}
\plottwo{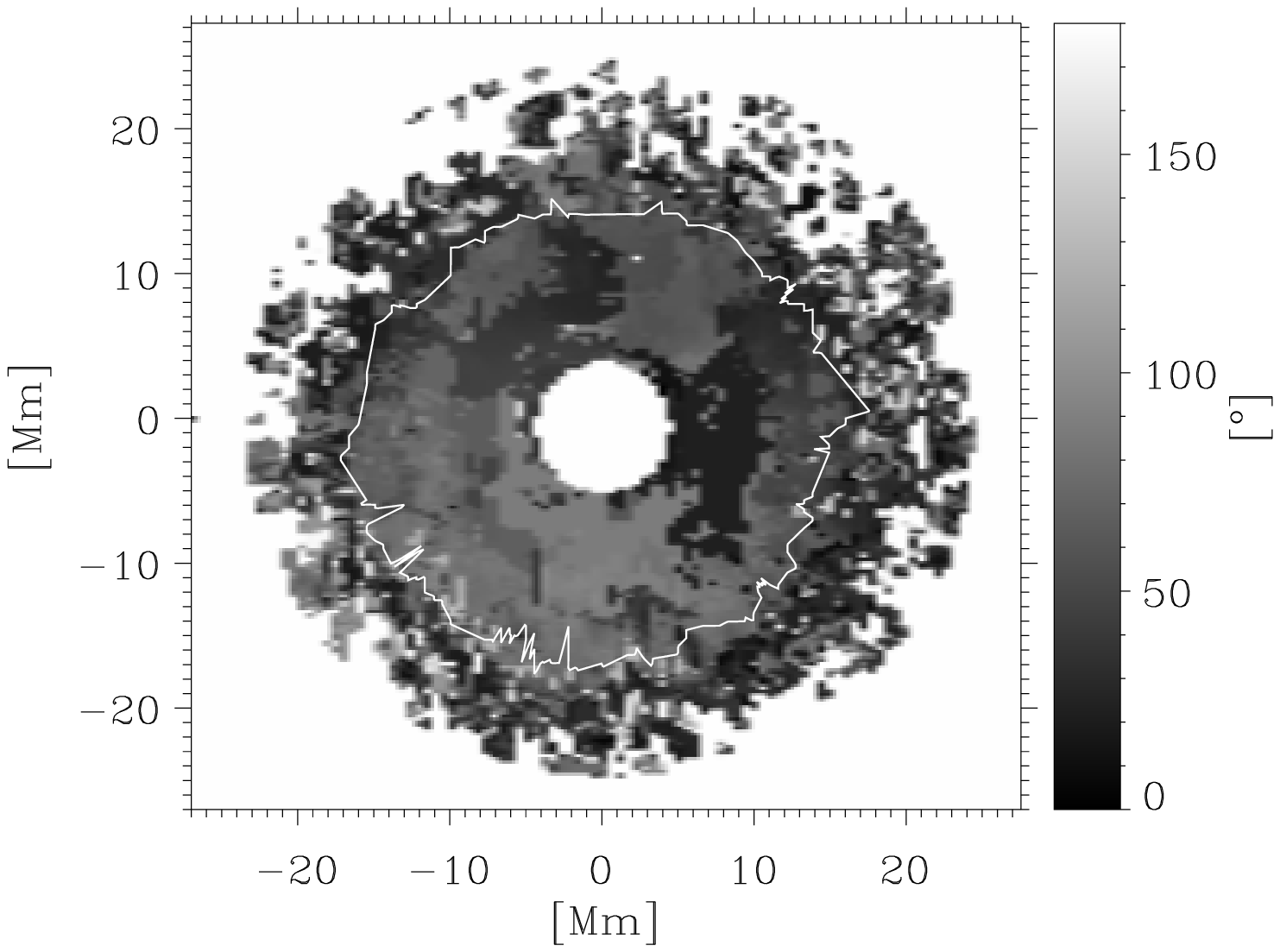}{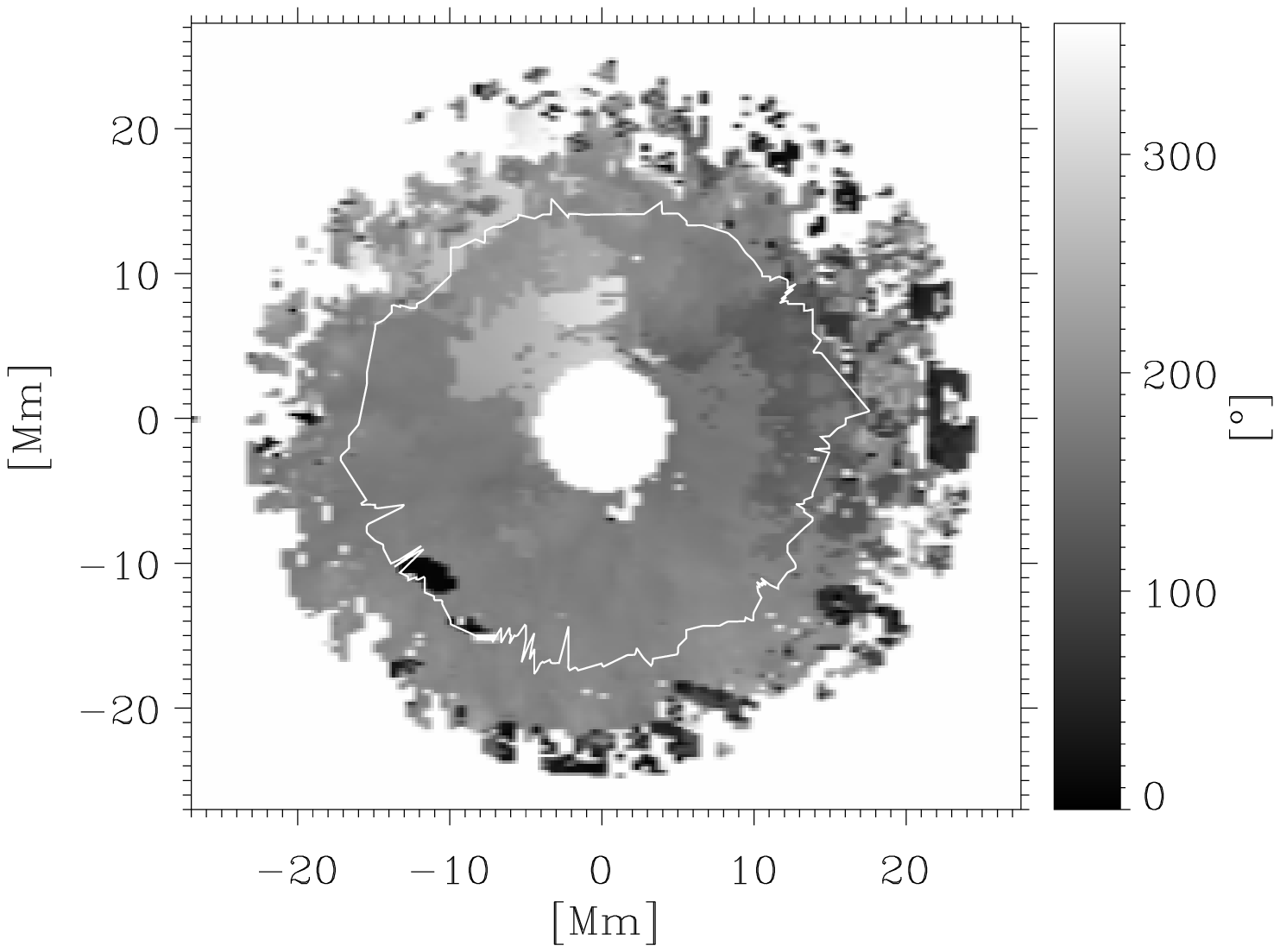}
\caption{Maps of the inclination (left) and the local azimuth (right) of the
minor magnetic component of the MISMA inversions.
They are represented as in Fig.~\ref{inc_az_image}.
The umbra is empty 
since the minor component is absent 
in this region.
}
\label{inc_az_image2}
\end{figure*}

	A rotation of the magnetic field vector
by 180$^\circ$ about the \los\ does not modify the Zeeman 
absorption \citep[e.g.,][]{unn56,lan92}. This leads to the
well known azimuthal ambiguity of the 
magnetic field derived from Zeeman magnetometry,
which has to sorted out before proceeding.
Since the MISMA model atmospheres have two magnetic
components (\S~\ref{scenario}), each one with this 180$^\circ$
ambiguity, each pixel has four different magnetic configurations
compatible with the 
observation\footnote{Two remarks are in order. First, neither 
the azimuth nor the
inclination of each component vary with
height in the atmosphere (\S~\ref{scenario}). 
Having varying azimuths and inclinations would make
the ambiguity intractable.
Second,
the Doppler shifts only depend on the  projection
of the velocity along the \los . Since 
we assume magnetic-field-aligned motions, all these
four solutions have the same velocity along the \los\ and
therefore are compatible with the inferred velocity field.
}.
Among them, we chose the one that best fits in 
a global magnetic field that
converges toward the sunspot core, and is  
as smooth as possible. First, we begin with the major components
of all pixels (i.e., with
the components containing most of the mass; see
\S\ref{scenario}). The smoothest solution is obtained
after an iterative procedure where the direction 
of the magnetic field vector in a given pixel 
is compared with the direction of the mean field
provided by the surrounding pixels. Then the 
solution closest to the mean direction is chosen.
This procedure is applied to each pixel and then
iterated to a point where no pixel over the full
FOV has to be modified. The iterative
procedure starts by choosing
that solution representing
a magnetic field converging toward the sunspot core
(see Fig.~\ref{schematic2}). If the two solutions of
a pixel are of this kind,
then the closest to the sunspot radial direction is chosen.
We tried various box sizes to compute the local average
(from 3$\times$3~pixels to 15$\times$15~pixels). The disambiguation
of the sunspot magnetic
field is almost insensitive to this parameter,
which mostly affects the field outside the visible
sunspot. We finally adopted 10$\times$10 pixels, which
provides a fair compromise.
Once the map of the major component is retrieved, we choose
that minor component solution whose
direction is closest to the major component direction.  
The algorithm sketched above provides a fairly axi-symmetric
magnetic field distribution centered at the sunspot core inferred
from the continuum intensity.
Figure \ref{inc_az_image} shows maps of inclination
and azimuth for the major components. 
Figure \ref{inc_az_image2} shows the same maps for the
minor components. 
%
%

%


\subsection{Setting a common scale of 
	heights for the full inversion}\label{scale_of_height}

Each pixel has been inverted independently from
the rest. Therefore each one has its own scale 
of heights. Part of the analysis of the forthcoming
sections becomes more  meaningful  when carried
out in a common scale of heights for the whole sunspot. 
We have set such common scale assuming
that the total pressure (gas pressure plus
magnetic pressure) is similar among neighbor pixels. 
(This approach was successfully 
applied in \paperii ; \S~4.2.)
The absolute vertical scale of each pixel is set by comparison
with the pressure of the nearest neighbors whose absolute scale
is already known. We minimize
the differences between the logarithms of the total
pressures allowing only for a global shift of the
scale of heights.
The procedure starts at the center of the 
sunspot and then continues to the penumbral
border, and beyond. 
The zero of the absolute scale is set 
by the 
layer $\tau_c=1$ in the stray light
component of the model atmospheres
far from the sunspot (further  than 23~Mm; see Fig.~\ref{cont_image}).
Theses model atmospheres represent the non-magnetic
Sun so that the absolute heights are referred to the base
of the non-magnetic photosphere.
The uncertainty introduced
by the assumption that the total pressure does not change
between adjacent pixels is discussed in Appendix \ref{appa}.
It is smaller than
$100$~km  for determining the 
Wilson depression of the umbra, and it becomes 
negligible small when computing pixel-to-pixel
variations ($\leq 4$~km).
%

\section{Main results of the MISMA inversion of a Sunspot}\label{main}

\subsection{Occupation fraction}\label{ocf_sect}
\begin{figure}
\plotone{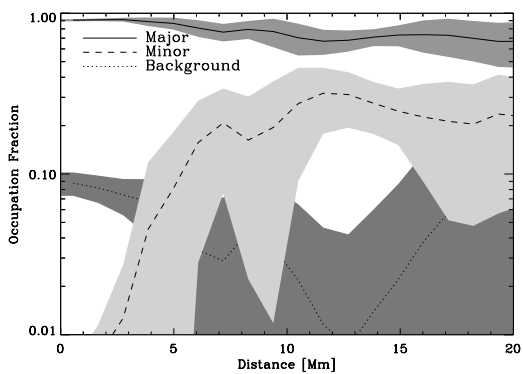}
\caption{Radial variation of the fraction of resolution element occupied
by each component of the model MISMAs.
The radial distance is given in Mm, and all occupation
fractions are evaluated at $\log\tau_c=-1$.
The shaded regions represent the standard deviation among the 
points at a distance from the sunspot core.
}
\label{ocf1}
\end{figure}

As described in \S~\ref{scenario}, the model MISMAs used for
inversion have three different components, two 
magnetic plus a non-magnetic  
background\footnote{The non-magnetic background should not be
confused with the stray light component included in the
fits (\S~\ref{scenario}, item~\#\ref{item5}). The stray light component
will not be discussed in the paper since
it shows a rather expected behavior similar to the non-magnetic stray light 
component deduced from the Milne-Eddington inversions of 
the same sunspot. The stray light fill factor goes from 2~\%
in the umbral core to 25~\% at the penumbral border.
}.
The {\em occupation
fraction} (OCF) is the fraction of atmospheric volume
occupied by each one of them. In principle
the OCFs
vary with height in the atmosphere (\paperi ) .
However, height variations are expected only
when the OCF of the background differs from zero, since
the magnetic OCFs grow at the expense of
non-magnetic volume. Since the OCF of the background
is very small (see below), the OCF of the magnetic 
components remains basically constant with height.
Figure \ref{ocf1} shows the retrieved OCFs
as a function of the radial distance from the sunspot center.
We plot the values at 
the height in the atmosphere where
$\log\tau_c=-1$, an optical depth 
where most of the atmospheric
information extracted from the inversion comes from.
The shaded areas represent plus and
minus one standard deviation among all the
points at the same radial distance.
Note that the major component has the largest OCF all the way from
the umbra to the sunspot surroundings. This result
is not a trivial consequence of the definition of 
major and minor, since the definition also
involves mass densities (\S~\ref{scenario}). 
The major component OCF goes from
more than 90\% in the umbra to about 70\% in the outer penumbra
(the solid line in Fig.~\ref{ocf1}).
The minor component is significantly different from
zero only outside the
umbra (from 5~Mm on; Fig.~\ref{ocf1}, the dashed line).
It occupies  30\% of the penumbral volume. 
%

%
%

%
\subsection{Magnetic field inclinations and azimuths}\label{sect_inc}
%
\begin{figure}
\includegraphics[angle=0,scale=.7]{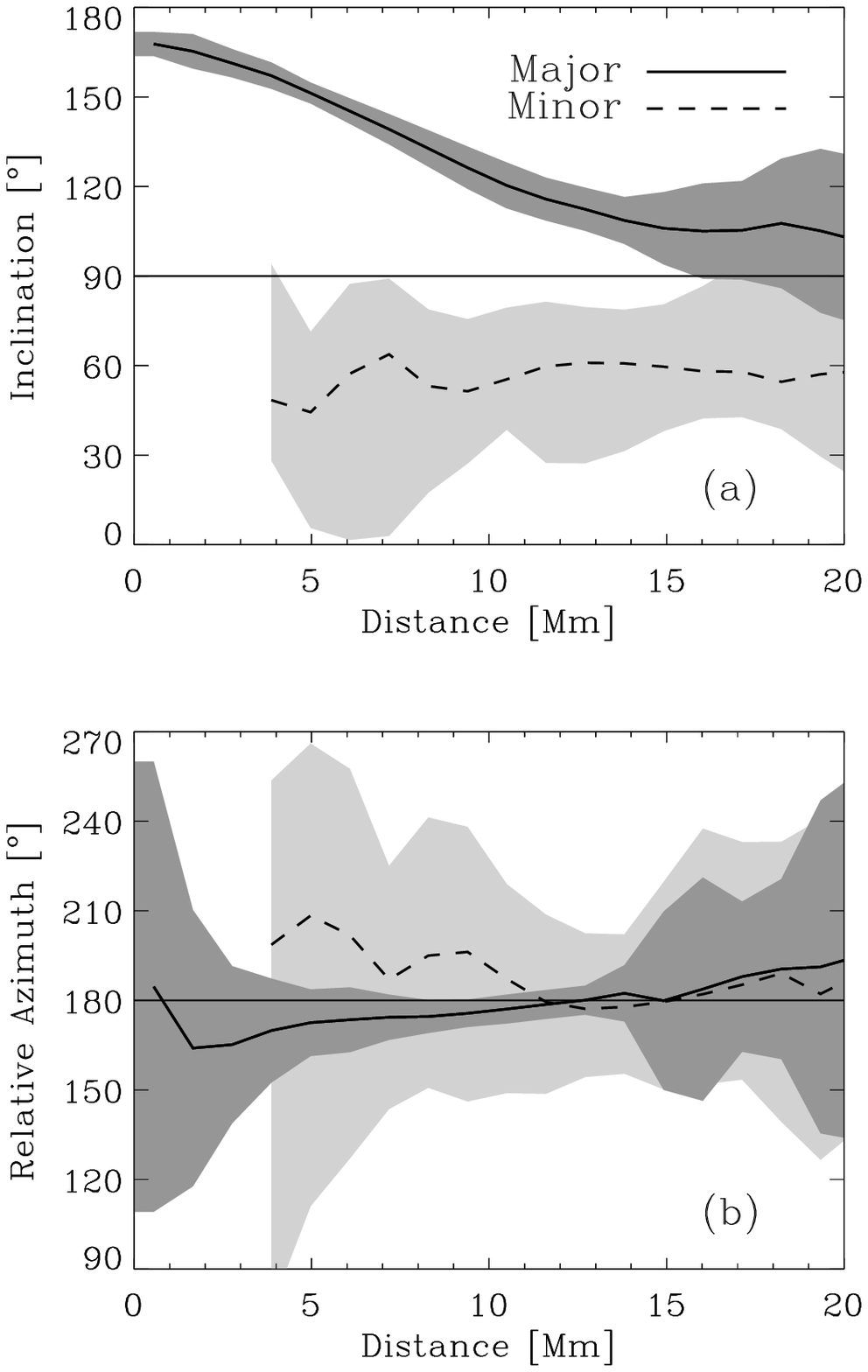}
\caption{Radial variation of the magnetic field inclination (a) and
relative azimuth (b) of the two magnetic components of the 
model MISMAs. 
The lines represent the mean values among those pixels
at a distance from the sunspot center. The shaded
regions correspond to the standard deviation.
Each type of line  identifies one
component, as described by the inset in (a).
}
\label{inc_az}
\end{figure}
\begin{figure}
\epsscale{1.}
\plotone{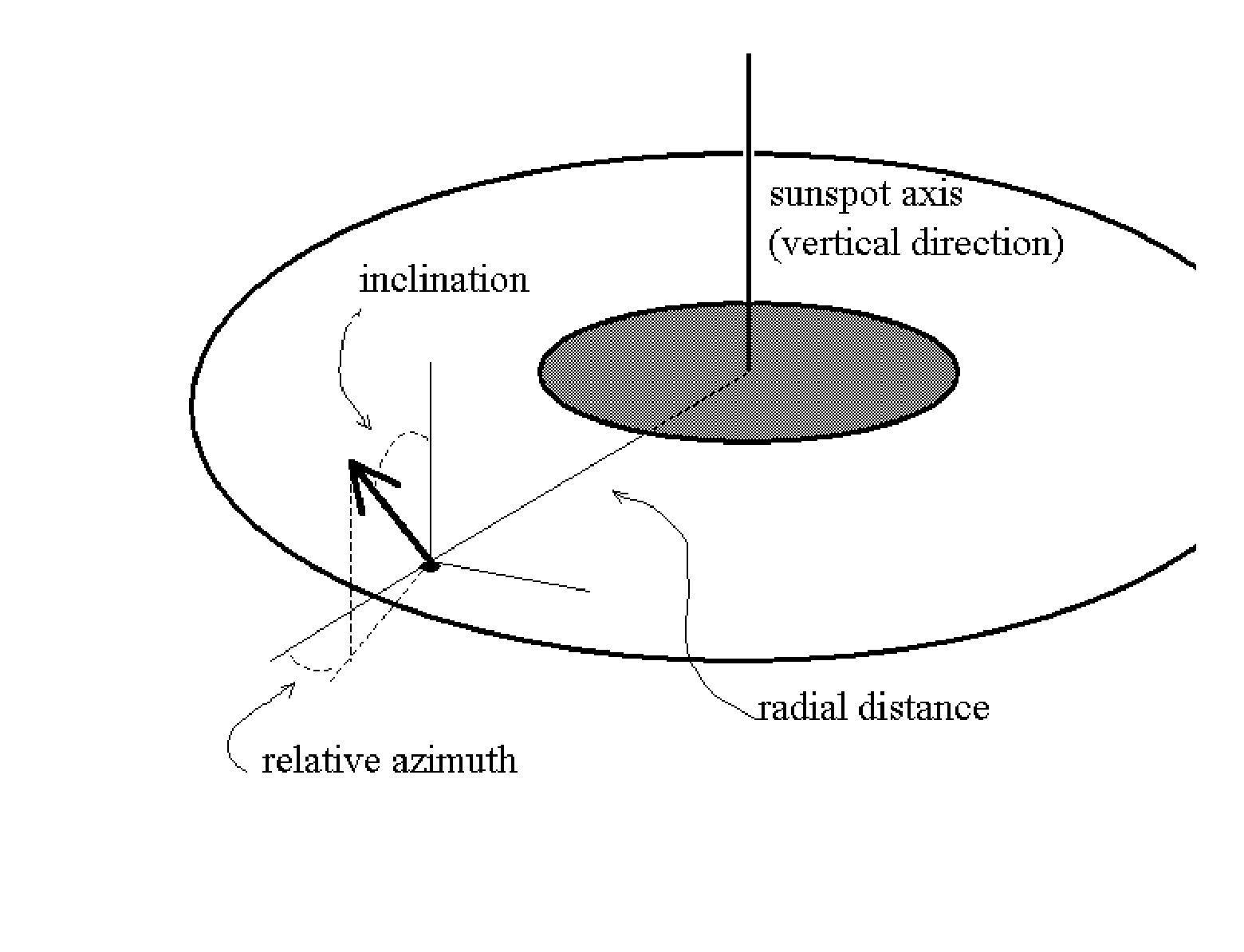}
\caption{Schematic to show the {\em inclination} and 
{\em relative azimuth} used to present the results 
of the inversion.
It displays the coordinate system used to describe
the magnetic field vector (the thick arrow) 
at a given radial distance from the sunspot core.
The relative azimuth is the azimuth in a local cylindrical
coordinate system where the radial direction is the radial
direction from the sunspot center. The inclination is
inclination with respect to the vertical direction. 
The coordinate system is different for different
pixels. For the sake of clarity, we show a magnetic
field of {\em positive} polarity, however,
the true sunspot is of {\em negative} polarity. For this 
reason the relative azimuths tend to be 180$^\circ$
(Fig.~\ref{inc_az}b).
}
\label{schematic1}
\end{figure}
\begin{figure}
\epsscale{1.}
\plotone{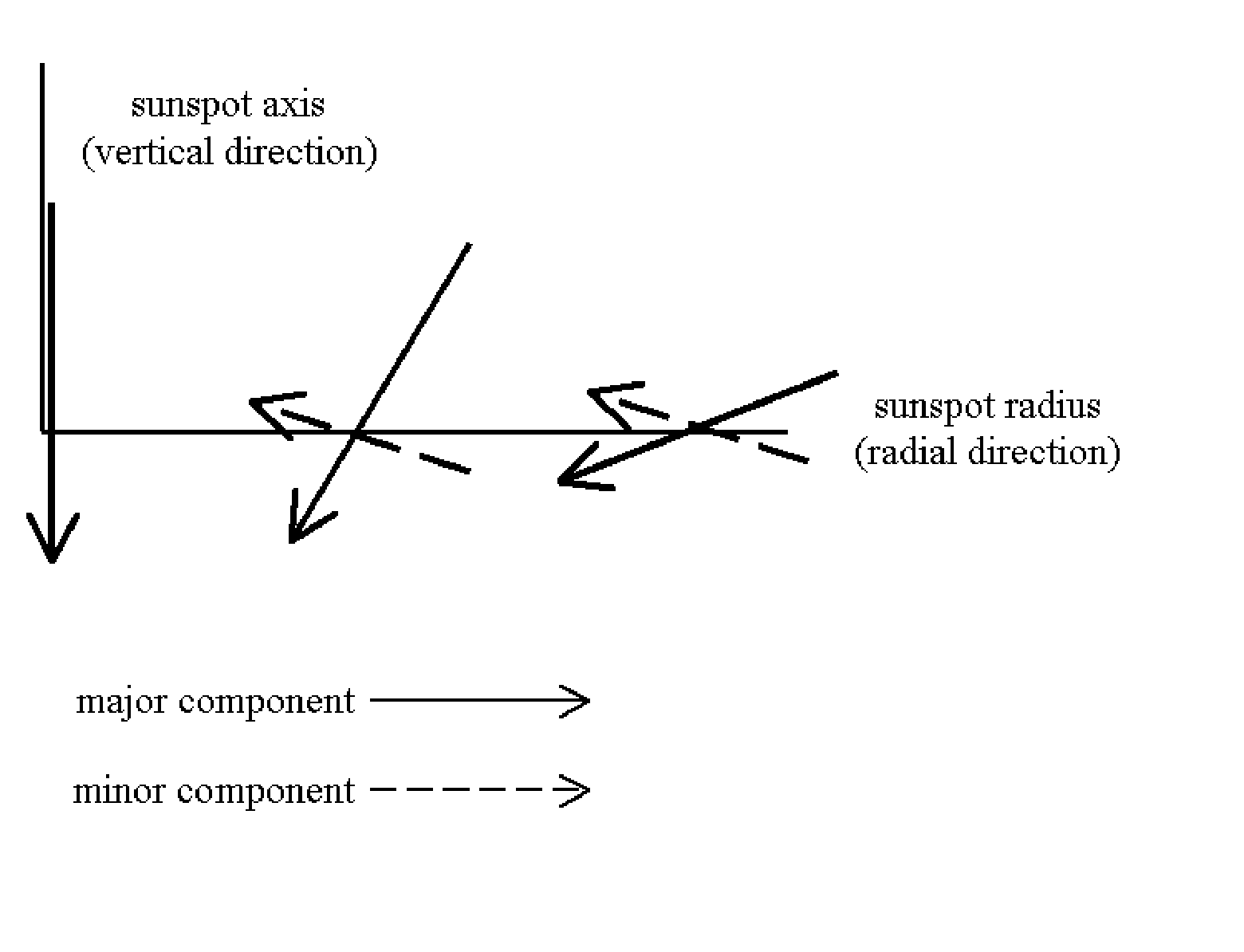}
\caption{
Schematic showing the direction
of the major and minor
components relative to the vertical and radial
directions. It shows the components
at three different radial distances. Since
we deal with a negative polarity sunspot, the magnetic
field points downward at the sunspot axis.
}
\label{schematic3}
\end{figure}

%
Figure \ref{inc_az_image} shows inclinations
and azimuths of the major component. 
Figure \ref{inc_az_image2} shows maps of the same
parameters for the minor component. 
The actual values 
are better appreciated in Fig. \ref{inc_az}, where
we represent the mean  
azimuths and inclinations, as well as the standard
deviation of these quantities among those
points at the same distance from the  sunspot center.
(For a visual definition  of vertical inclination and 
relative azimuth, see Fig. \ref{schematic1}.)
Both the major component and (to less extent) the minor
component follow the radial direction. The
relative azimuth with respect to the radial direction
is of the order of 180$^\circ$, as it should be 
for a radial magnetic field in a negative polarity
sunspot (Fig. \ref{inc_az}b). (The properties
of the minor component for distances 
smaller than  5~Mm are excluded.
The minor component of these points does not 
contribute to the observed signals and therefore,
any inference based on it results unreliable.)
The magnetic field inclination of the major component
follows a variation with radial
position typical of sunspots 
(from say, 170$^\circ$ at the sunspot core, 
to about 15$^\circ$ off the horizontal at the 
sunspot boundary; see., e.g., Fig.~9 in \citealt{lit93}). 
Together with this
smooth and expected behavior, the minor component is
(with large scatter but beyond any doubt) always
less inclined than the horizontal
(see the dashed line in  Fig.~\ref{inc_az}a). 
We deal with a negative polarity
sunspot and, therefore,  visualizing the  magnetic
topology from the inclination is not so
straightforward.
The cartoon in Fig.~\ref{schematic3}
has been laid out to illustrate the magnetic
topology.
This inclination of the minor
component is common to the whole penumbra, implying
that magnetic field lines are returning to the solar
interior over the entire penumbra. 
This is possibly the biggest surprise of these MISMA
inversions. 
As we will discuss in connection with the mass
flows (\S~\ref{evershed}), inferences of magnetic fields returning
to the sub-photosphere  at the outer
penumbra and beyond do exist in the literature
\citep{rim95,wes97}.
The presence of magnetic field lines that 
return 
throughout the penumbra is new, though.
%
%
%

%
\subsection{Magnetic field strengths and fluxes}\label{mfs}
\begin{figure}
\plotone{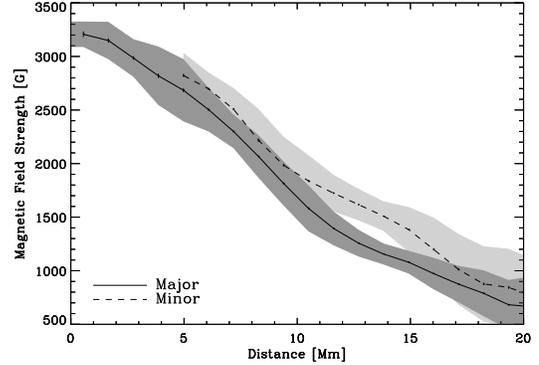}
\caption{Radial variation of the magnetic field
strength of the two magnetic components. 
The mean field strength of the minor component is  10\% larger
than the strength of the major component.
The shaded regions represent the standard deviation among all the
points at a distance from the sunspot core. The lines correspond
to the mean values. Error bars for these mean values are also
included, but they go unnoticed due to their small values.
}
\label{mfs_fig}
\end{figure}

Figure \ref{mfs_fig} shows the variation of the magnetic field
strength with the distance from the sunspot center. 
The field strength varies with height, so we
plot the value at $\log\tau_c=-1$.
The two magnetic components
are shown separately. (The field strength of the minor
component in the umbra is not included since its occupation
fraction is negligible and therefore its value dubious;
\S~\ref{ocf_sect}.)
The field strengths
go from 3~kG at the sunspot core to 1~kG at the
transition between penumbra and its surroundings.
The observed variation is typical
of sunspots 
\citep[e.g., Fig.~9 in][]{lit93}.
More interesting is the systematic difference
of field strength between components.
The mean field strength
of the minor component is about 200~G larger
than that of the major component. 
This excess is significant. The dispersion among the 
field strengths of all the pixels at a 
radial distance is also of the order of 200~G (the
shaded regions in Fig.~\ref{mfs_fig}). The dispersion
of the mean 
is reduced by a factor
square root of the
number of pixels contributing to each radial point.
Typically, several hundred pixels contribute to the mean
so that the uncertainty of the mean is only a few tens of
G, and so, well below the observed 200~G difference. 
The error bars of the mean are also included in
Fig.~\ref{mfs_fig}, but they go almost unnoticed due to
their small values.

%
%
%

%
The excess of field strength creates a surplus of
magnetic pressure and, consequently, a deficit of gas
pressure in the minor component.
Let us assume that 
the minor and major components of the same or
neighbor resolution elements are connected
by field lines. (We have not been able to 
come up with any other possibility to understand the 
presence of field lines pointing up and down
throughout the penumbra. However, this connection between the
field lines of the minor and major components does
not result from the inversions and
is speculative.)
Then
the gradient of gas pressure between minor and
major components is bound to drive
a mass flow. This is nothing but the siphon flow mechanism invoked 
to explain the Evershed flow  
\citep{mey68,tho93}.
The seemingly small difference of field strength
that we infer
can drive a very fast flow,
as attested by a simple order of magnitude estimate.
For a typical magnetic field
strength in the penumbra $B$ of  2~kG,
\begin{equation}
\Delta B/B\sim 0.1.
\end{equation}
The square velocity difference produced by a gas pressure
excess $\Delta P_g$ is 
\begin{equation}
\Delta U^2\sim -\frac{2\Delta P_g}{\rho},
	\label{equili}
\end{equation}
where we have used the equation of motion
for a horizontal thin flux tube  \citep[e.g.,][]{spr81}.
Since the gas pressure difference is due 
to the magnetic field strength difference,  then
\begin{equation}
\Delta P_g=-\frac{B^2}{4\pi}\frac{\Delta B}{B}.
	\label{equili2}
\end{equation}
Using equations~(\ref{equili}) and (\ref{equili2}),
one finds,
\begin{equation}
(\Delta U^2)^{1/2}\sim \Big[\frac{B^2}{2\pi\rho}\frac{\Delta B}{B}\Big]^{1/2},
\end{equation}
with the symbol $\rho$ standing for the density.
For typical densities at $\log\tau_c\sim-1$ ($\sim 10^{-7}$~g~cm$^{-3}$;
see \S~\ref{sdensity}),
\begin{equation}
(\Delta U^2)^{1/2}\sim 8~\rm{km~s}^{-1}.
\end{equation}
This velocity is similar to the minor component
velocities inferred
from the inversions; see \S~\ref{evershed}.

\begin{figure}
\includegraphics[angle=0,scale=.7]{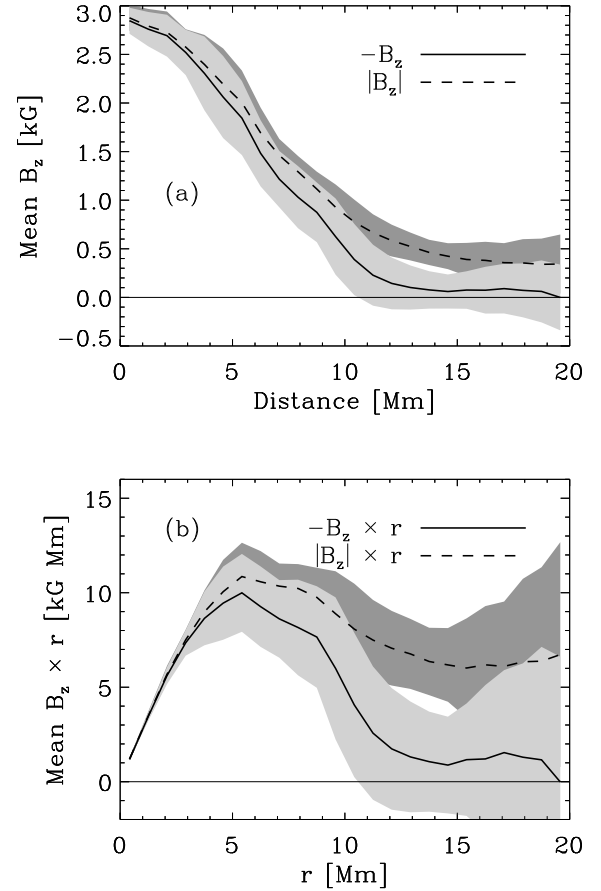}
\caption{(a) Volume average of the vertical magnetic field
strength. The solid line shows the signed mean
whereas the dashed line corresponds to the unsigned
mean. Note the large difference in the 
external penumbra (say, for distances 
between 10 and 16~Mm.)
The shaded  regions represent plus-minus
one standard deviation among all those points at the 
same distance from the sunspot center.
(b) Average vertical field times distance to the
sunspot center ($r\equiv$~Distance). The integral
of this quantity between two radii
provides the vertical magnetic flux.
}
\label{net_flux}
\end{figure}

According to the current paradigm, a large fraction
of the sunspot magnetic flux emerges
in the penumbra \citep[e.g.][]{sol93c}. 
It leads to the concept of deep penumbra,
where the boundary between 
magnetized plasmas and non-magnetic environment
occurs well below the observed photospheric layers.
To our surprise, the model sunspot 
turns out to have
very little vertical magnetic flux across 
the penumbra. Our finding does not  imply
a shallow penumbra, though. It follows from
the large amount of return flux in the penumbra.
The two magnetic components  tend to have
vertical magnetic field strengths with
opposite signs (\S~\ref{sect_inc} and Figs.~\ref{inc_az} and
\ref{schematic3}).
This leads to a cancellation of the 
average vertical component of the field
strength $<B_z>$. 
(The angle brackets $<~>$ are used in the paper
to represent volume average, i.e., an
average weighted with the OCFs.)
The mean vertical component at $\log\tau_c=-1$ is shown in  
Fig.~\ref{net_flux}a.
It tends to zero in the external penumbra
(say, for distances between 10~Mm and 16~Mm).  
However, the unsigned mean $<|B_z|>$ is certainly different
from zero.
Such difference between signed mean and unsigned mean
produces significant differences between 
the signed and unsigned magnetic fluxes of the
sunspot.
The signed vertical magnetic flux
of those points with radii between 
$r_1$ and $r_2$ is
\begin{equation}
\Phi_z(r_1< r < r_2)=2\pi\int_{r_1}^{r_2}r\widehat{<B_z>}
	dr,
	\label{def_flux}
\end{equation}
with $\widehat{<B_z>}$ the azimuthal average
of $<B_z>$,
\begin{equation}
	\widehat{<B_z>}={{1}\over{2\pi}}\int_0^{2\pi}
		<B_z>d\chi,
\end{equation}
and $\chi$ the absolute azimuth of each point.
(The unsigned mean is formally identical to
equation [\ref{def_flux}] replacing $<B_z>$ with $<|B_z|>$.)
Figure~\ref{net_flux}b shows the integrand of 
equation~(\ref{def_flux}), i.e.,
the contribution of each radius
to the sunspot flux. It is clear how the
penumbra makes a significant contribution to the
unsigned flux of the sunspot, but little to the
signed flux.  Table~\ref{table1} quantifies how much.
It lists the vertical magnetic flux of the sunspot, the umbra and
the penumbra.
The signed flux of the sunspot is  33\% smaller
than the unsigned flux. Flux cancellation primarily
occurs in the penumbra (58\% reduction) and very little
in the umbra (9\% reduction). The cancellation
at the penumbral border 
is more than 80\%,
a result which could be naturally explained if
field lines emerge, bend over, 
and return 
in a distance similar to the angular resolution.
(A detailled discussion about the physical 
interpretation of such speculative arguments 
is deferred to another paper; \citealt{san04c}.)
The penumbra has as much unsigned
flux as the umbra, but only 40\% of the
signed flux.
Table~\ref{table1} also includes the fluxes of a
standard model sunspot without return flux.
The flux of the standard model sunspot lies in between 
our signed and unsigned fluxes. 
%

%
\begin{deluxetable*}{lllcc}
\tablewidth{0pt}
\tablecaption{Vertical magnetic flux across the sunspot} 
\tablehead{\colhead{}&\colhead{}&\colhead{Unsigned}&
	\colhead{Signed}&\colhead{Reference\tablenotemark{a}}}
\tablenotetext{a}{Normalized sunspot model in \citet{sol00}, Table~14.34,
	with a maximum vertical field strength of 2.8~kG and 
	the radius of our sunspot.}
\tablenotetext{b}{We assume the umbral radius to be
	half of the sunspot radius $R_S$.}
\tablenotetext{c}{Penumbral rim having half
	of the sunspot area.}
\startdata
$\Phi_z(0<r<R_S)$&full sunspot&$\Phi_0=7.3\times 10^{21}$\phd 
Mx&0.67\phd$\Phi_0$&
	0.99\phd$\Phi_0$\\
$\Phi_z(R_S/2<r<R_S)$&penumbra\tablenotemark{b}&
	$\Phi_1=0.49\phd\Phi_0$&0.42\phd$\Phi_1$&
	0.83\phd$\Phi_1$\\
$\Phi_z(R_S/\sqrt{2}<r<R_S)$&outside 
penumbra\tablenotemark{c}&$\Phi_2=0.25\phd\Phi_0$&
	0.20\phd$\Phi_2$&0.63$\phd\Phi_2$\\
$\Phi_z(0<r<R_S/2)$&umbra&$\Phi_3=0.51\phd\Phi_0$&
	0.91\phd$\Phi_3$&1.14\phd$\Phi_3$\\
\enddata
\label{table1}
\end{deluxetable*}

%
\subsection{Mass Density}\label{sdensity}
\begin{figure}
\plotone{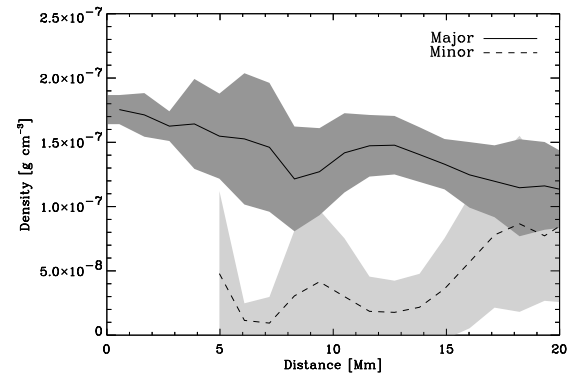}
\caption{Radial variation of the density 
of the two magnetic components. 
The major component is  five times denser than
the minor component.
Distances are given in Mm from the sunspot center.
}
\label{density}
\end{figure}

Figure \ref{density} shows the variation of the density
along 
the radial direction at $\log\tau_c=-1$. The densities of the
major component are 
not very different from the
densities of the quiet Sun at this optical depth. Note, however,
how the density of the minor component is five times smaller.
The association of minor component with low density
is partly due to the definition
of minor and major, which involves the density. In addition,
the enhanced magnetic field strength of the minor component
described above demands a decrease of gas pressure and,
consequently, a decrease of density. (Recall that
major and minor 
temperatures are forced to be identical; \S~\ref{scenario}.)
%
%

%
\subsection{Motions and the Evershed effect}\label{evershed}
All motions are aligned along the magnetic field lines.
This is a constraint directly imposed on
the model MISMAs (\S~\ref{scenario}).

\begin{figure*}
\epsscale{.9}
\includegraphics[angle=90,scale=.7]{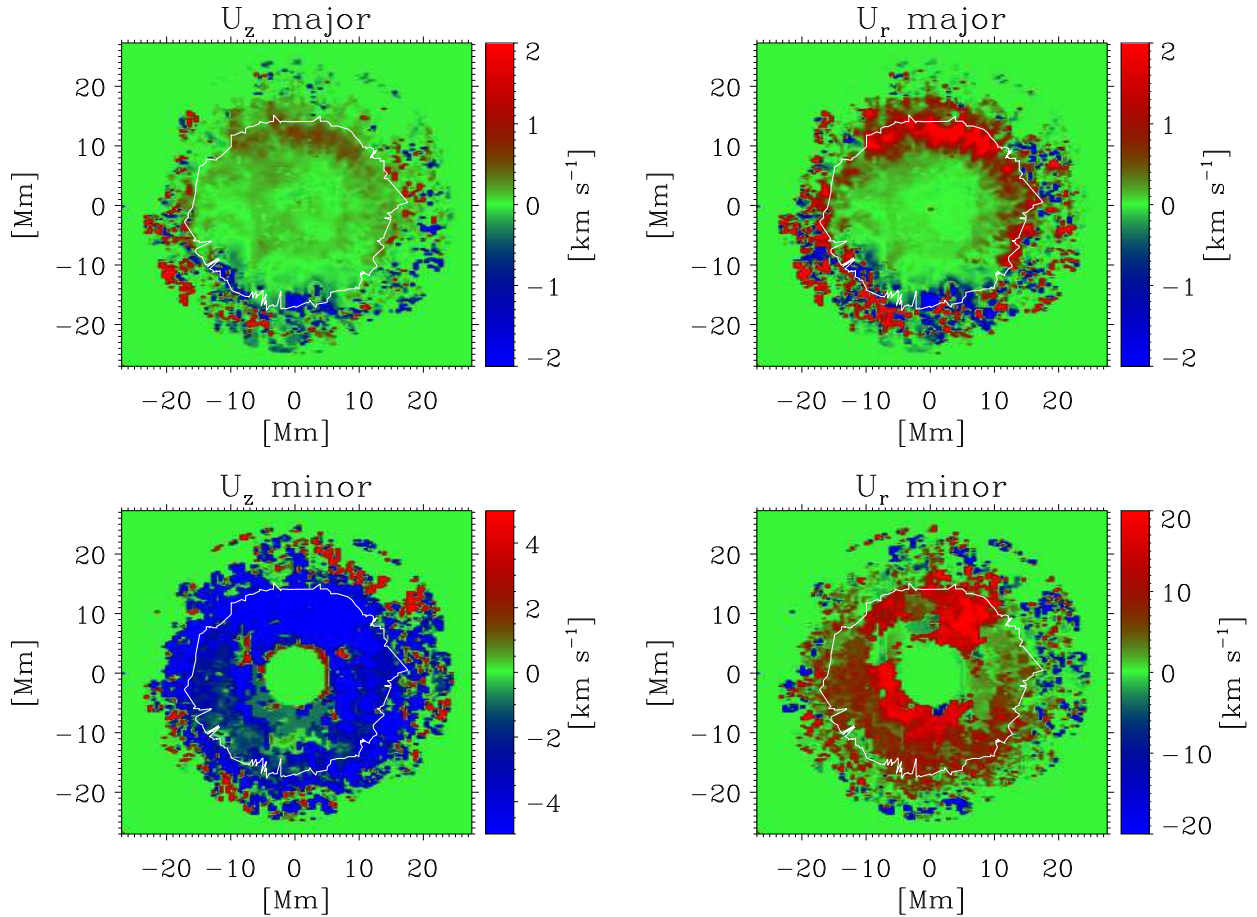}
\caption{Horizontal $U_r$ and vertical $U_z$ velocities
of the two magnetic components at $\log\tau_c=-1$.
The equivalence between the color code and the velocity
differs for the different plots, as it is indicated by the
vertical bars. 
Positive vertical velocities correspond to upward motions. 
Positive radial velocities correspond to outward motions.
First, note the large velocities of the minor component
(the two figures at the bottom), 
as compared to the velocities of the major component
(the two figures at the top).
Second, note the downward motions of the minor component
plasma occurring over the entire penumbra.
The minor component signals  have been set to zero
in the umbra,
since the minor component  does not contribute to the
umbral spectrum.
The white contour outlines the border of the
continuum sunspot,
i.e., the transition between the penumbra and
the surrounding photosphere (c.f. Fig.~\ref{cont_image}).
}
\label{evershed2b}
\end{figure*}

\begin{figure}
\includegraphics[angle=0,scale=.7]{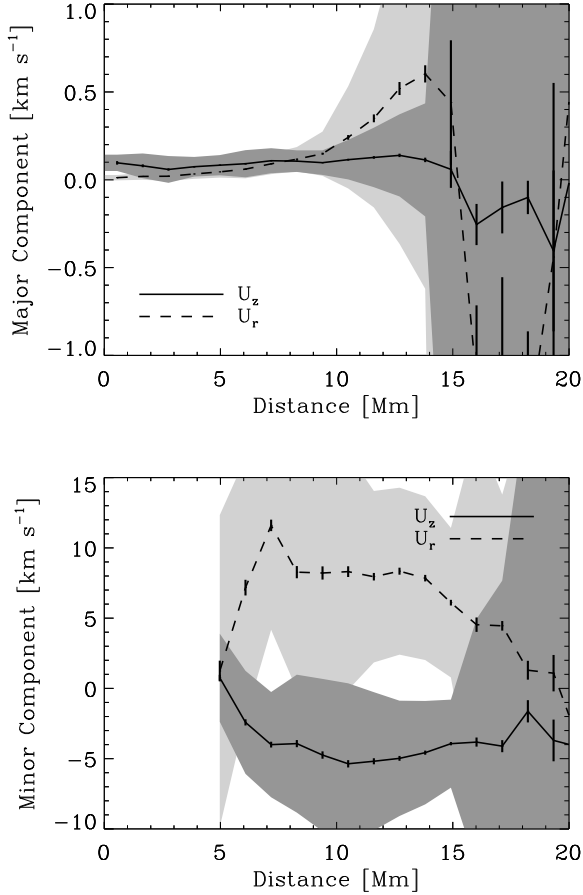}
\caption{Horizontal $U_r$ and vertical $U_z$ 
velocities of the two magnetic components
of the model MISMAs at $\log\tau_c=-1$.
We represent the mean (lines) and the standard  deviation
(shaded regions) for all pixels  at a 
distance from the sunspot center. 
The major component is in the upper plot whereas
the bottom plot corresponds to the minor component.
Positive velocities are upward ($U_z$) and
outward ($U_r$).
The vertical bars correspond to the formal error bars
of the mean velocities.
}
\label{evershed3a}
\end{figure}

The velocity pattern retrieved from the inversion is simple and
regular; see  Fig.~\ref{evershed2b}.
The minor component, with the  
lowest mass density and the highest magnetic field strength,
has very large velocities. They typically reach 10 km~s$^{-1}$ at
$\log\tau_c=-1$. The major component
shows much smaller values. Both 
velocities are directed radially outward 
along the field lines.
Since  the magnetic field lines are not horizontal
(see, Fig.~\ref{inc_az} and \S~\ref{sect_inc}) the velocities
have both horizontal and vertical components.
Figure~\ref{evershed3a} shows the radial velocity $U_r$  and vertical
velocity $U_z$ of the major (top) and minor (bottom)
magnetic components at $\log\tau_c=-1$. The vertical velocity
of the major component is positive, meaning that 
it is directed upward. The vertical velocity
of the minor component is directed downward. This
downward directed velocity occurs all over the
penumbra (Fig.~\ref{evershed2b}).  
The small upward motions in the umbra (Fig.~\ref{evershed3a}, top)
are consistent with the lack of motion found by \citet{bec77}
considering the uncertainties of the absolute scale of velocities
(\S~\ref{observations}).
%
%

%
\begin{figure}
\includegraphics[angle=0,scale=.7]{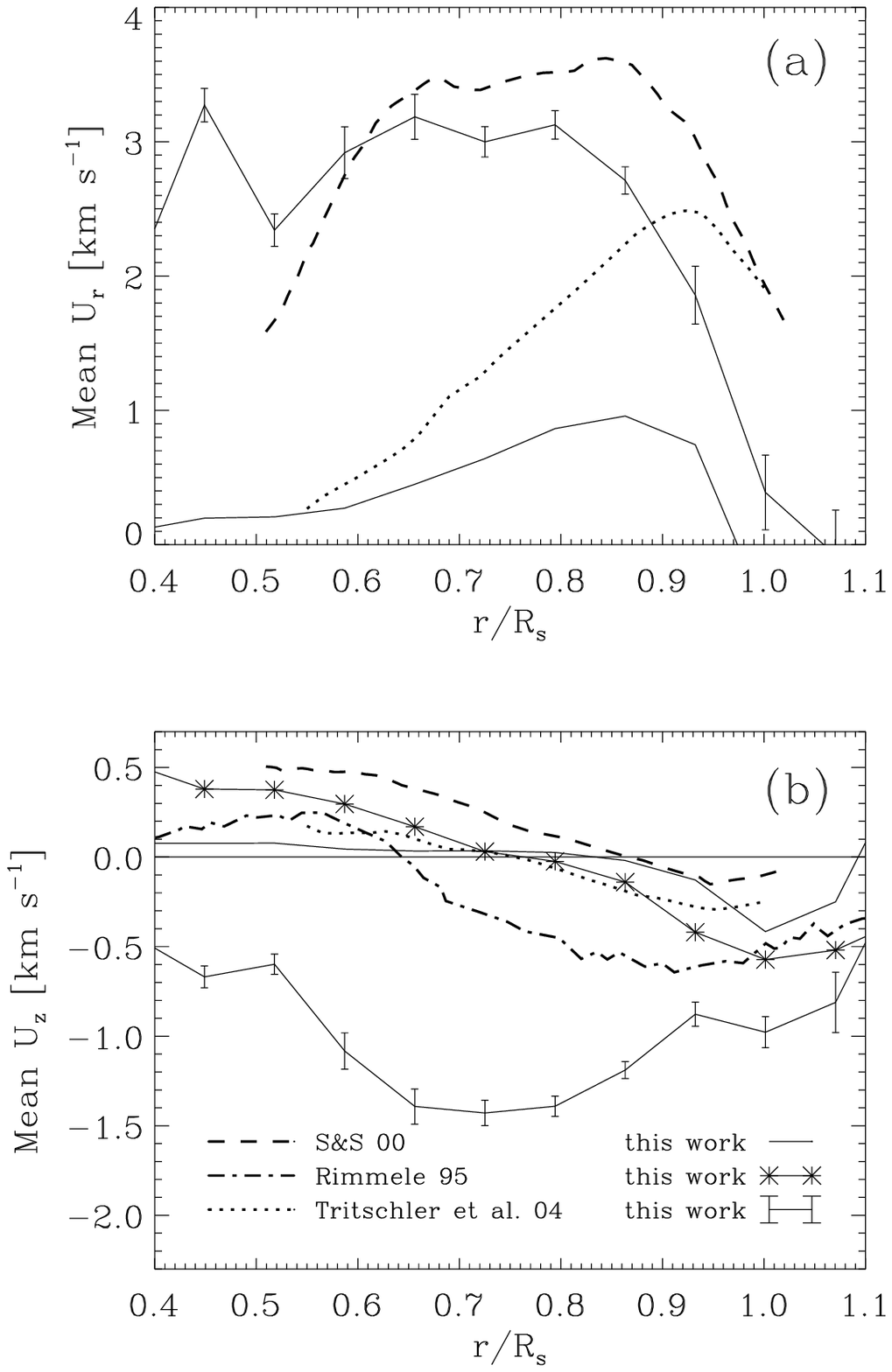}
\caption{Average horizontal (a) and vertical (b) 
velocities as retrieved from the MISMA inversions.
Positive means radially outward (a) and vertically
upward (b). The solid lines with error bars show the 
volume average of the two components,
whereas the plain solid lines correspond to  
the density weighted
average.
The other lines are shown for reference, and
they correspond to works finding
downflows within the sunspot boundary (\citealt{rim95}, the
dotted dashed line;
\citealt{sch00}, the dashed line, where we only
represent one among the four  targets;
\citealt{tri04}, the dotted line).
The solid line with asterisks shows the mean vertical
velocities we deduce from the bisector of \linb , simulating
one of the measurements from the literature.  
The radial distance $r$ has been referred to the
sunspot radius $R_S$ to allow comparison between
sunspots of different radii.
}
\label{evershed4}
\end{figure}

%
The finding of upward motions in the inner penumbra and
downward motions in the outer penumbra
is now common in the recent literature 
\citep{rim95,sch00,del01,bel03b,tri04}.
To the best of our knowledge, we are
the first to infer ubiquitous 
downward motions in a penumbra.
(Recall, ubiquitous upward and 
downward motions.)
This seemingly inconsistency of our
finding with previous works is probably
an artifact produced by 
the spatially unresolved  structure of the flows.
Most techniques employed so far 
assume uniform velocities in the resolution element.
However, observations show with insistence
the presence of several velocity components
\citep{bum60,ste71,wie95,rue99}. In our case
we infer two very different types of 
motion\footnote{The finding of two types
is an idealization that we impose, and
it should be understood
as the need for more than one mean component
to describe the observed line shapes.}.
When the observations are interpreted as a single
resolved component, the {\em measured}
velocity corresponds to some kind of ill-defined mean value
of the actual  
velocities\footnote{Like
the well-known convective blueshift
of the photospheric
lines \citep[e.g., ][]{dra82}.
If this convective blueshift is interpreted as a 
net motion, one infers a significant global expansion
solar surface, able to double the solar
radius in less than a month.
}.
Since (a) the true velocities can be
extremely high (e.g., Fig.~\ref{evershed3a}), 
(b) there are both upward and downward motions, and
(c) the residual velocities are very small,
{\em measuring} upflows or downflows depends 
on subtleties of how the various motions contribute
to the mean values.
The contribution depends on the physical
properties of the penumbral atmospheres, as well
as on method employed to measure (the spectral line,
the portion of line profile that is
employed, the
method used 
to assign Doppler shifts, and so on).
In order to show that this interpretation 
of the systematic flows deduced by others 
is consistent with our inversions,
we have also measured the Evershed  flow
of our sunspot with one of the techniques used in
single component inferences.
Downflows at the penumbral rim have been observed 
using the displacement of the Stokes $I$ bisector at the
line wings \citep[e.g.,][]{rim95,tri04}. 
We compute the bisector corresponding to each observed
Stokes $I$ profile of \linb , the line 
with lowest Zeeman sensitivity in the spectra.
The azimuthally averaged shifts of the bisectors 
are represented in Fig.~\ref{evershed4}b 
(the solid line with asterisks).
(The azimuthal average of the \los\ velocity
cancels the contribution 
of the horizontal motions, so that only
vertical velocities remain; see, e.g.,
\citealt{sch00}.) 
The simulated bisector-based measurement
shows upflows in the inner penumbra and downflows
in the outer penumbra, in agreement with  single component
estimates (several representative measurements from the
literature are  included in Fig.~\ref{evershed4}).
The bisector-based mean velocity is qualitatively 
different from the true mean velocity
of our inversions (the solid lines with error
bars in Fig.~\ref{evershed4}). 
This OCF weighted mean velocity
shows downflows everywhere, meaning that
the minor component dominates the average.
However, the nature of the curve changes
when the mean is computed weighting each
component with its mass (i.e., mass density
times OCF). Then the vertical velocity  
is not far from 
to some of the mean velocities of the literature
(compare the plain solid line in Fig.~\ref{evershed4}b
with, e.g., the dotted line).
One may think that the scatter among individual
velocities in Fig.~\ref{evershed3a} questions  
the small mean velocities shown in
Fig.~\ref{evershed4}. However, small differences
of mean velocities are significant
since hundreds of independent inversions contribute to 
each mean value. The scatter of the mean 
decreases as 
the square root of the number of pixels contributing
to the mean, a factor of the order of 30 at the penumbral
edge. The plain mean velocities in Fig.~\ref{evershed4} include
these formal error bars, supporting the reliability
of our mean velocities. 
(The errors for the mass weighted mean
are similar.)
Formal error bars for the mean velocities are also included
in Fig.~\ref{evershed3a}.

%
%

%
\subsection{Net Circular Polarization}\label{ncp_sect}
\begin{figure}
\epsscale{1.0}
\plotone{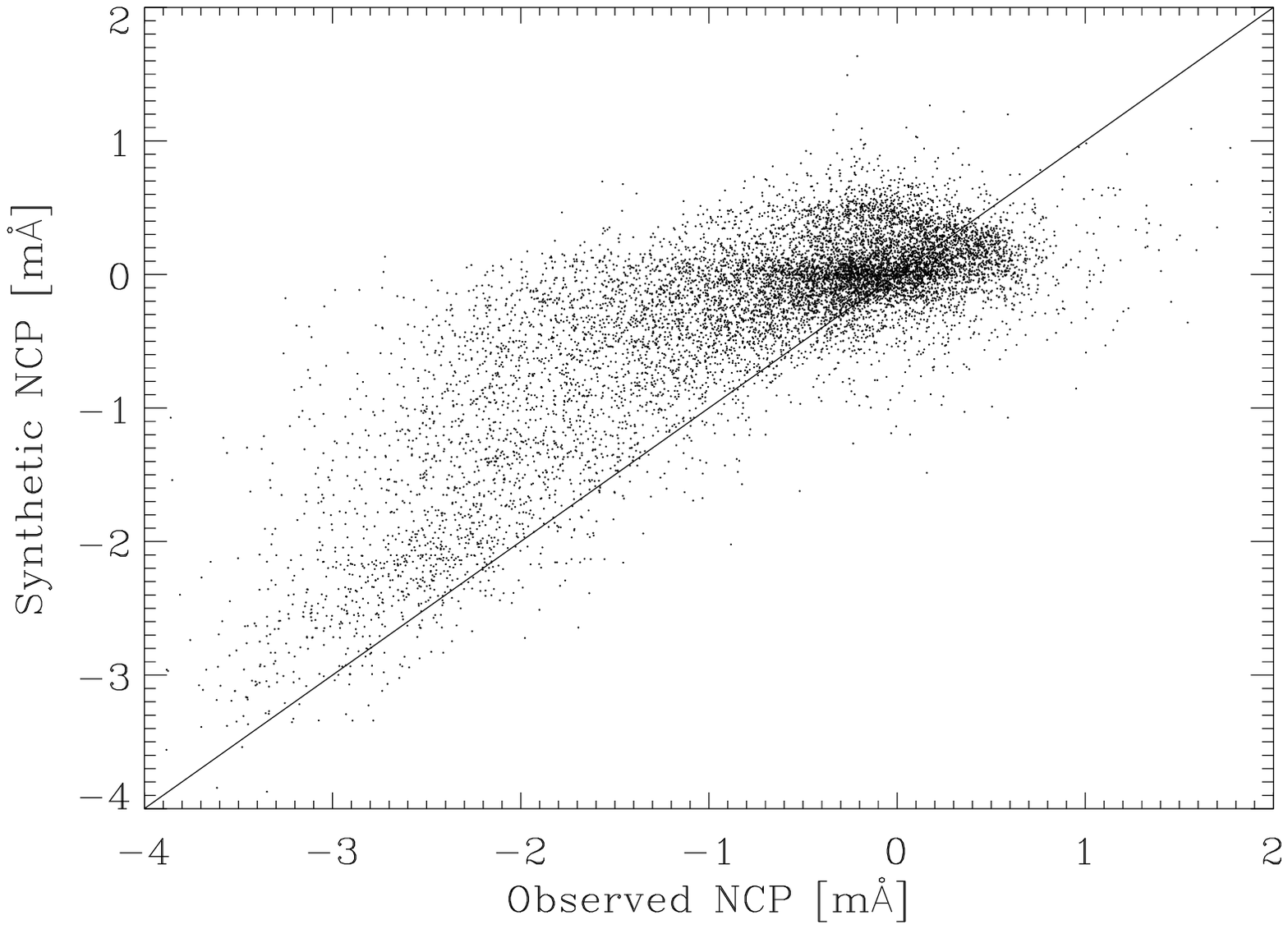}
\caption{Net Circular Polarization (NCP) of \lina . 
Scatter plot of synthetic NCP versus 
observed NCP. The synthetic values 
slightly underestimate the observed ones
(the solid line corresponds to equal synthetic and
observed NCPs). However,
observations and modeling
follow the same tendency and have the same spatial
distribution across the sunspot (see main text).
}
\label{ncp}
\end{figure}

The presence and behavior of Net Circular Polarization (NCP)
created by individual spectral lines has stimulated some of  
the important advances on the penumbral magnetic field
structure. Any consistent penumbral model should be able
to reproduce its properties. Our semi-empirical
sunspot does it. The NCP is defined as  
\begin{equation}
{\rm NCP}=\int_{\lambda_1}^{\lambda_2} V(\lambda)d\lambda,
\end{equation}
with the integral extending a full line profile
included between wavelengths $\lambda_1$ and $\lambda_2$.
Figure~\ref{ncp} shows a scatter plot of synthetic versus 
observed NCP of \lina . Although the observed values tend 
to be larger than the synthetic ones,  
the syntheses grasp the observed behavior.
The maximum value is around -3~m\AA\  and  the signs
are preserved. In addition, both the observed  NCP
and the synthetic NCP 
follow the general rules found by
\citet{ill74a,ill74b} and \citet{mak86}, and
summarized in \citet[\S~4]{san92b};
the extreme NCP is negative (as the sunspot polarity)
and occurs at the limb-side penumbra where
the Stokes $V$ profiles show the cross-over effect
(profiles with three or more lobes like  those in Fig.~\ref{show_fit}a).
%

\begin{figure}
\epsscale{1.1}
\plotone{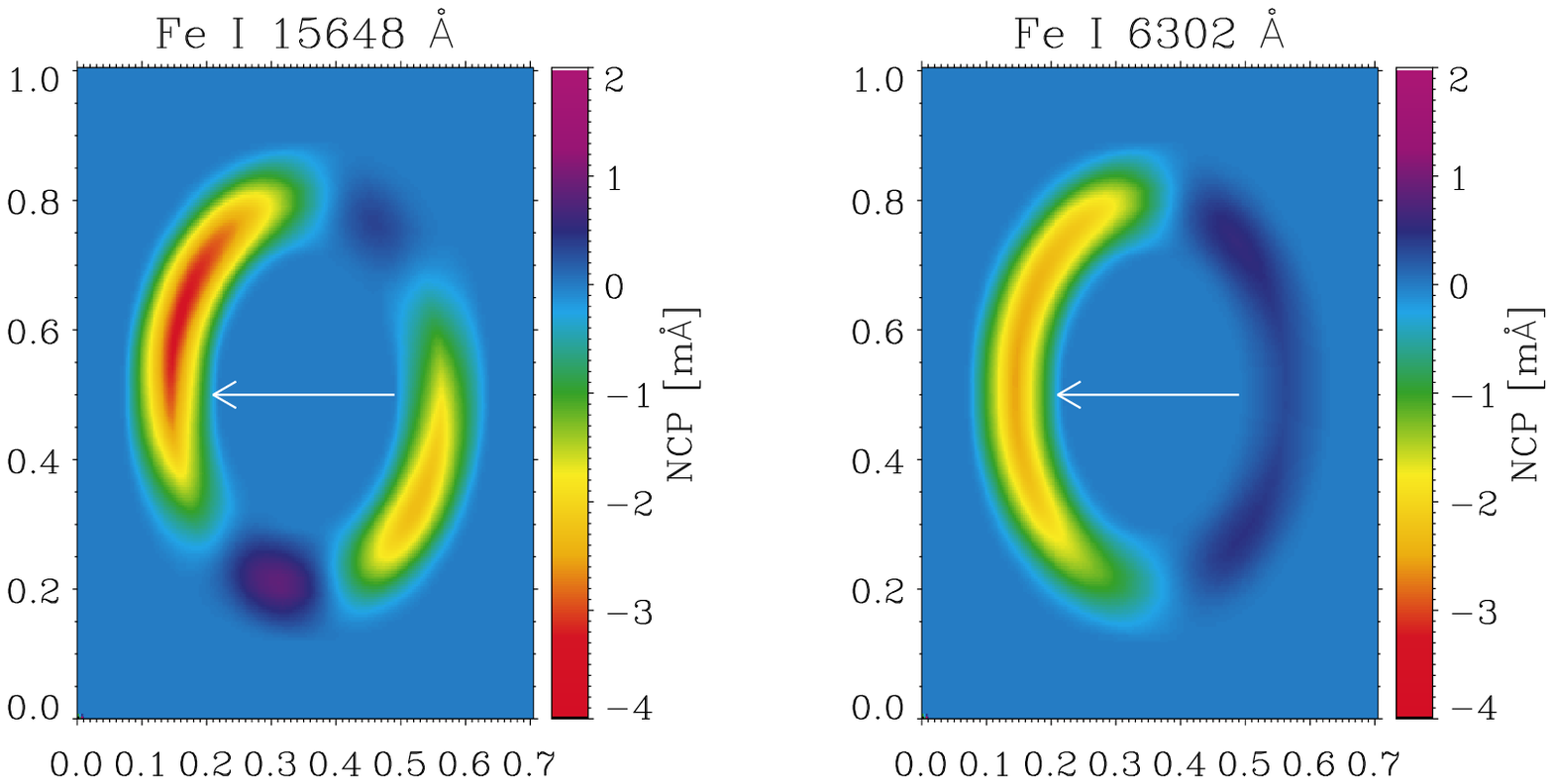}
\caption{Net Circular Polarization (NCP) of \linc\ (left)
and \lina\ (right) when one of the model MISMAs is used
to produce an axi-symmetric magnetic field distribution.
This axi-symmetric distribution is observed from the side
to mimic a sunspot observed at $\mu=0.7$. Note how the
NCP of \lina\ is almost symmetric with respect to the line 
along the solar radial direction (indicated by the arrow,
whose tip points the solar limb direction).
This symmetry disappears for \linc , an effect
observed by \citet{sch02b}, and explained by 
\citet{sch02c} and \citet{mul02} in terms of a thick 
horizontal fluxtube embedded in a more vertical
magnetic background. Our models also
explain this observation. See text for details.
The spatial coordinates are given in arbitrary 
units. 
}
\label{ncp3}
\end{figure}

In order to add one more piece of evidence supporting
the consistency of the model sunspot,
we synthesized the NCP produced by \linc.
The NCP of this line is peculiar since
it does not obey the rules of the
NCP mentioned in the previous paragraph,
in particular, the NCP is not symmetric with respect 
to the line dividing the
sunspot along a solar radial direction
\citep[][ Fig. 8, \S~4.3]{sch02b}.
Due to this symmetry breaking, the extreme NCP for \linc\ 
does not occur in the penumbral regions closest
to the solar limb. This behavior is
attributed by \citet{sch02c} and \citet{mul02} to the action
of Magneto-Optical Effects (MOE) in the radiative
transfer. 
MOE are significant only for 
very large Zeeman splittings \citep[e.g.,][]{lan73},
which is indeed the regime to be applied to
the strongly Zeeman sensitive
\linc\ line. 
Due to the existence gradients of magnetic field
azimuth along the \los , MOE create
NCP out of linear polarization signals
\citep{sch02c,mul02}. These gradients of azimuth 
are produced
by the existence of two magnetic components in the
resolution element having
different magnetic field inclination with
respect to the vertical. 
When the \los\ is not along the vertical direction,
the differences of vertical
inclination render differences of azimuth in the plane perpendicular
to the \los .
(For a very intuitive representation of the effect,
see Fig.~4 in \citealt{kal91}.)
The magnetic geometry required  by \citet{sch02c} and
\citet{mul02} 
is already present in the model sunspot that we retrieve 
and,
in principle, it has the potential to reproduce the
observed \linc\ NCP.
Figure~\ref{ncp3} shows the NCP produced
by \linc\  assuming  an  axi-symmetric sunspot  
located at $\mu=0.7$. (We have used one randomly 
chosen  model MISMA to represent the thermodynamic
state
and the velocities in the atmosphere of this axi-symmetric
sunspot.) 
The synthetic \linc\ NCP turns out to be in good agreement with
the observed spatial distribution.
Figure~\ref{ncp3} also includes the NCP of \lina , 
which is symmetric with respect 
to the line along the solar radial direction
(arrows in Fig. \ref{ncp3}).
%

%
\subsection{Wilson depression}\label{sect_wilson}

\begin{figure}
\includegraphics[angle=0,scale=.4]{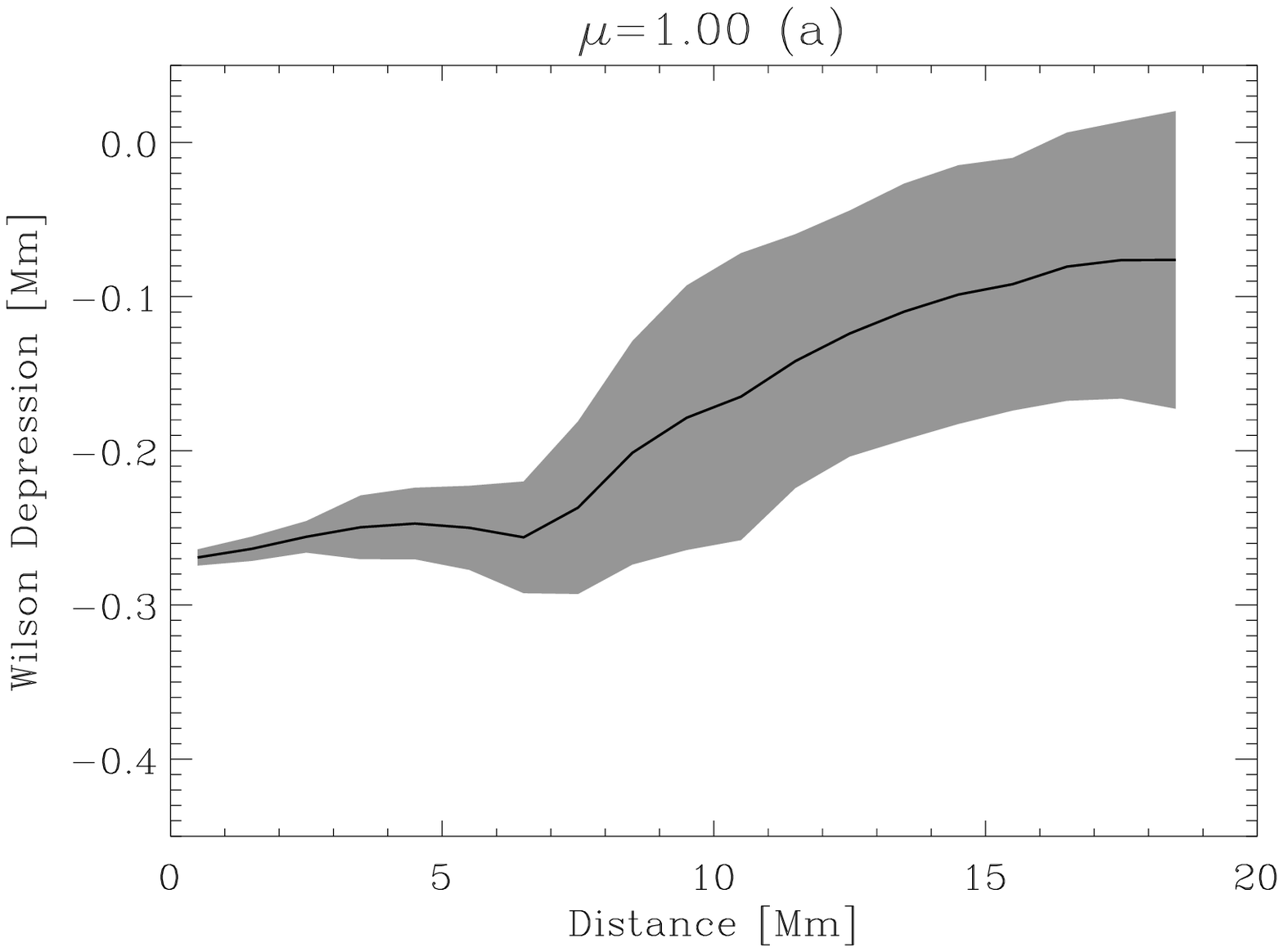}
\includegraphics[angle=0,scale=.4]{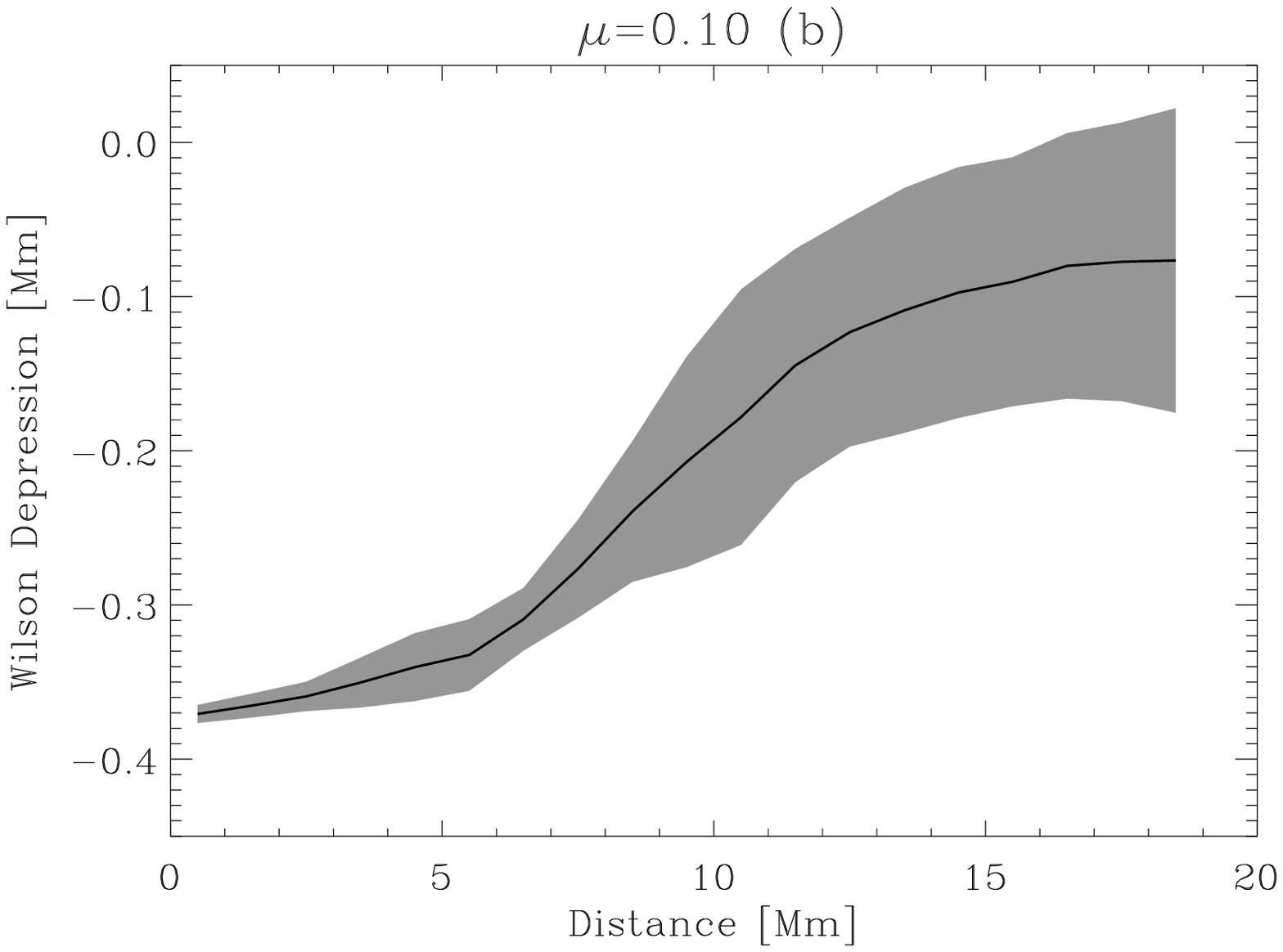}
\caption{Wilson depression as a function of the radial distance
	within the sunspot. The shaded region corresponds
	to the standard deviation among those points
	having the same radial distance from the sunspot center.
	(a) Wilson depression for the observed sunspot,
	with $\mu\simeq 1$.
	(b) Wilson depression 
	if the sunspot were close to the solar limb ($\mu=0.1$).
	Note how the Wilson depression increases
	toward the limb. 
	}
\label{wilson0}
\end{figure}

The presence of a strong magnetic field
evacuates and cools down the sunspot atmosphere with
respect to the non-magnetic quiet Sun. The
resulting drop of density
produces a depression of the 
observed sunspot photosphere
with respect to the
non-magnetic photosphere.
Figure~\ref{wilson0} shows the depth of the 
layer $\tau_c=1$ as a function of
the position in the sunspot (i.e.,
the Wilson depression). Refer to \S~\ref{scale_of_height} 
for the estimate  of absolute heights.
The Wilson depression of the sunspot core
is smaller than
the values reported in the literature,
even keeping in mind the 100~km uncertainty
worked out in Appendix~\ref{appa}. For example,
{\em Allen's Astrophysical Quantities} 
quotes (-0.6$\pm$0.2)~Mm \citep{sol00}, whereas we
get $-0.25$~Mm instead (Fig.~\ref{wilson0}a). 
Despite the discrepancy, we trust our estimate 
because the semi-empirical model sunspot reproduces  the 
observations on which the large Wilson
depressions reported in the literature
are based on. 
%

\begin{figure}
\epsscale{1.0}
\plotone{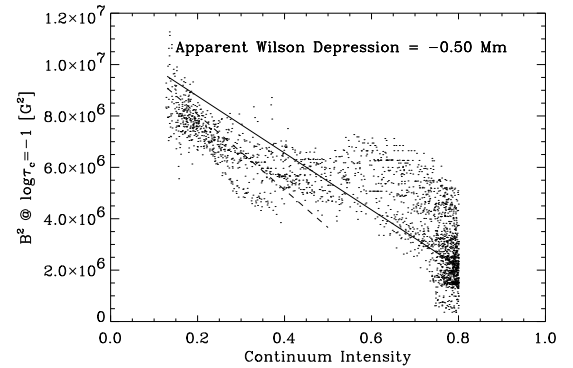}
\caption{
Scatter plot of the square of the magnetic field strength
$B^2$
versus continuum intensity. 
Linear fits of this empirical relationship have been
used to estimate the Wilson depression
of sunspots at the solar disk center. The two 
linear regressions 
consider continuum intensities~$< 0.8$ (the solid line) and 
continuum intensities~$< 0.5$ (the dashed line). Both provide an
apparent Wilson depression of about -0.5~Mm,
which overestimates the true Wilson depression
of the model sunspot (-0.25~Mm).
}
\label{wilson_wrong_f}
\end{figure}

%
First, there are estimates based on the 
correlation between magnetic field strength $B$ and
continuum intensity $I_c$ \citep[e.g.,][]{kop92,mar93}:
\begin{equation}
I_c=\Big(1-\frac{B^2}{4\pi P(W_D)}\Big)I_{c0}.
	\label{wilson_wrong}
\end{equation}
The symbol $P(W_D)$ stands for the gas pressure at the Wilson
depression in the non-magnetic Sun, where $I_c=I_{c0}$. 
The relationship (\ref{wilson_wrong}) is inspired by the 
condition of magnetostatic equilibrium of an axi-symmetric
sunspot (for details, see the papers cited above).
The Wilson depression $W_D$ is determined carrying out
a linear regression $B^2$ versus $I_c$ that provides a slope $m$
and a constant $k$,
\begin{equation}
I=mB^2+k.
\end{equation}
Then, inverting equation~(\ref{wilson_wrong}),
\begin{equation}
W_D=P^{-1}\big[-k/(4\pi m)\big],
	\label{wilson_wrong2}
\end{equation}
where the depth corresponding to a given gas pressure
(the inverse $P$ function or $P^{-1}$)
is derived from \citet{spr74} convection zone model.
Figure \ref{wilson_wrong_f} shows the scatter plot between
the observed continuum intensity and the mean magnetic
field strength provided by the inversion at an optical
depth representative of the formation of the spectral
lines ($\log\tau_c=-1$). It also shows two linear fits
obtained using $I_c < 0.8$ (the solid line) and 
$I_c < 0.5$ (the dashed line). The application of 
equation~(\ref{wilson_wrong2}) 
renders $W_D\simeq~-0.5$~Mm in both cases.
This figure is consistent with the values 
reported in the literature, but overestimates the
true Wilson depression of the model sunspot
by a factor two. 
%
%

\begin{figure}
\epsscale{1.}
\plotone{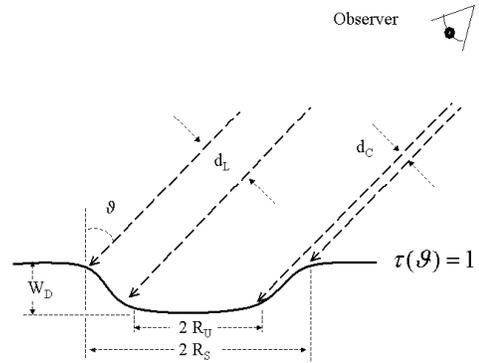}
\caption{Schematic of a sunspot observed away from
the solar disk center.
The solid line represents the height in the atmosphere where
the continuum optical depth equals one at the heliocentric angle
of observation $\theta$. The figure also shows
various symbols used to describe the Wilson
effect of our model sunspot, in particular, the
different foreshortening of the limb-side penumbra $d_l$
and the center-side penumbra $d_c$.
}
\label{wilson1}
\end{figure}

\begin{figure}
\epsscale{1.}
\plotone{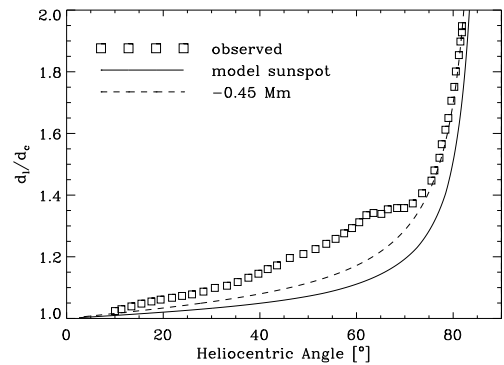}
\caption{
Ratio between the size of the limb-side penumbra $d_l$
and the
center-side
penumbra $d_c$ versus heliocentric angle of the sunspot. 
The crude estimate from our model sunspot (the
solid line) provides a fair description of the mean
observation (the symbols). The dashed line is a best fit
description of the observations at the limb,
and it yields a Wilson depression
of -0.45~Mm. Heliocentric angles are given in degrees.
}
\label{wilson2}
\end{figure}

%
Second, most Wilson depression measurements have been carried
out for sunspots at the solar limb. Close to the limb the
Wilson effect becomes
conspicuous, and it allows estimating the depression from simple
geometrical arguments. In particular, the center-side penumbra 
becomes shorter 
than the limb-side penumbra \citep[e.g.,][ \S 3.8]{bra74}.
The estimates of the Wilson depression
based on this foreshortening also exceed the value 
that we infer at disk center. However, the Wilson depression 
of our model
sunspot varies with the position on the disk. It increases
toward the solar limb. The physical basis is simple:
pressure and density scale heights are proportional to  
the temperature and, therefore, they are much
smaller in the umbra than in the non-magnetic photosphere. 
Consequently, the range of atmospheric heights contributing 
to the observed
light is narrower in the umbra. 
The umbral photosphere is not only deeper but thinner.
Since limb observations reveal high photospheric
layers, the Wilson depression necessarily increases toward the limb.
We have estimated the center-to-limb variation
of the Wilson depression 
as follows.
For plane parallel atmospheres, 
the optical depth of an atmospheric layer 
changes with the position on the disk as
\begin{equation}
\tau_c(\mu)=\tau_c(1)/\mu,
	\label{tau_at_mu}
\end{equation}
with $\mu$ the cosine of the heliocentric angle $\theta$,
and $\tau_c(1)$ the optical depth at disk center.
The Wilson depression at $\mu$ is
the height in the atmosphere having $\tau_c(\mu)=1$, which 
corresponds to the disk center optical depth
equals $\mu$. In general, the plane parallel approximation is not valid 
to describe our  model sunspot close to the limb, since the photon mean
free path becomes larger than the pixel size at the disk center.
However, it is a fair approximation to describe the Wilson depression
of the umbral core, where
the physical properties vary 
on a scale much larger than the resolution element (e.g., the Wilson
depression is rather uniform in the central 5~Mm, see Fig. \ref{wilson0}a).
The same approximation is also valid in
the quiet Sun, which we have 
used to set the absolute height of the Wilson depression
(\S~\ref{scale_of_height}). Figure \ref{wilson0}b
shows the Wilson depression at $\mu=0.1$ 
($\theta\simeq 84^{\rm o}$) following from
equation~(\ref{tau_at_mu}). 
It turns out to be larger than the depression at the disk center,
explicitly, close to $-0.4$ Mm in the umbral core. 
Is this value compatible
with the observations of the Wilson effect near the limb? 
It provides a fair description of the most conspicuous property
of the Wilson effect, namely, the differential foreshortening of the
limb-side penumbra $d_l$ as referred to the center-side penumbra
$d_c$ (see Fig. \ref{wilson1}). Using simple geometrical arguments, 
it is possible
to write down a expression for this foreshortening that depends
only on the depression of the umbra $W_D$ and the radii of the
umbra $R_U$ and the penumbra $R_S$, 
\begin{equation}
d_l/d_c={
	{(R_S-R_U)\mu+W_D\sqrt{1-\mu^2}}
	\over
	{(R_S-R_U)\mu-W_D\sqrt{1-\mu^2}}
	}.
	\label{foreshortening}
\end{equation} 
(Expressions similar to equation~[\ref{foreshortening}] are used to infer
Wilson depressions from observed foreshortenings,
e.g., \citealt{chi62}.)
Using 
equation~(\ref{foreshortening}) and the Wilson depressions of 
the model umbra at various
heliocentric angles, we compute the differential foreshortening.
It is shown in Fig. \ref{wilson2}. The same figure includes values 
reported in the literature 
(the square symbols labeled as {\em observed}; 
\citealt{jen69}).
The model sunspot  follows
the observed tendency. The ratio is a little too small, but
keep in mind that the observed curve represents mean values, and the
individual sunspots do show significant deviations from these 
mean values (see., e.g., the points in Fig. 6 of \citealt{jen69}).
If one fits the foreshortening observed at the limb using
as a constant free parameter
$W_D$ in  equation~(\ref{foreshortening}),
$W_D\simeq-0.45~$Mm 
(see the dashed line in Fig. \ref{wilson2}). 
This value is close to that in
Fig.~\ref{wilson0}b, which 
allows us to argue that the low Wilson depression of our model
sunspot at $\mu=1$ is consistent with the Wilson depressions
reported for sunspots at the limb ($\mu\ll 1$).
%
%
%

Note yet another property of the Wilson depression 
in our model sunspot. The outer penumbra
is depressed with respect to the quiet 
photosphere by about 100~km. The curves in
Fig.~\ref{wilson0} do not show a sharp transition
at the penumbral boundary (distance $\sim$~17~Mm)
because they represent the Wilson depression of the  
magnetized fraction of the 
atmosphere. However, outside the penumbral border the 
stray light contamination of the model atmospheres 
occupies most of the resolution elements,
and the Wilson depression of this stray light is
basically zero. Consequently, there is a 
significant discontinuity of the observed layers
at the penumbral border, which is
bound to complicate the interpretation of all
observables.

To sum up, the model sunspot 
obtained from the inversion provides large  apparent
Wilson depressions consistent with those reported
in the literature,
despite the fact that the
true Wilson depression is rather small. 
In addition, there is a significant discontinuity
of the depression
at the penumbral border.
%

\section{Magnetic flux conservation}\label{mfc_sect}
	
	One can reproduce the asymmetries of the
Stokes profiles observed in sunspots assuming that the
magnetic field varies monotonically with height 
\citep{san92b,wes01a,mat03}. It requires atmospheres having 
vertical gradients of magnetic field inclination.
However, the gradients are too large to fit in 
the global topology of the sunspot magnetic field
\citep{san92b}. \citet{sol93a} 
formulate the problem 
pointing out that the divergence of the
resulting magnetic field differs from zero and, consequently,
the magnetic flux is not conserved. 
A way out was proposed by \citet{sol93b}. The 
gradient of magnetic field inclination does not have 
to be monotonic. 
A horizontal magnetic field embedded in a more inclined
background field also provides the observed Stokes
asymmetries. \citet{sol93b} use a single thick
horizontal fluxtube to explain the idea, but
a collection of thin fluxtubes does the same
effect. The latter
is actually the MISMA model proposed by \citet{san96},
and the approach followed in this paper
(\S~\ref{scenario}). Since the
conservation of magnetic flux is central 
to assess the
consistency of the retrieved model atmospheres,
we have checked it with care.
We use the differential formulation $\nabla{\bf B}=0$,
with {\bf B} the magnetic field vector.

\begin{figure}
\includegraphics[angle=0,scale=.85]{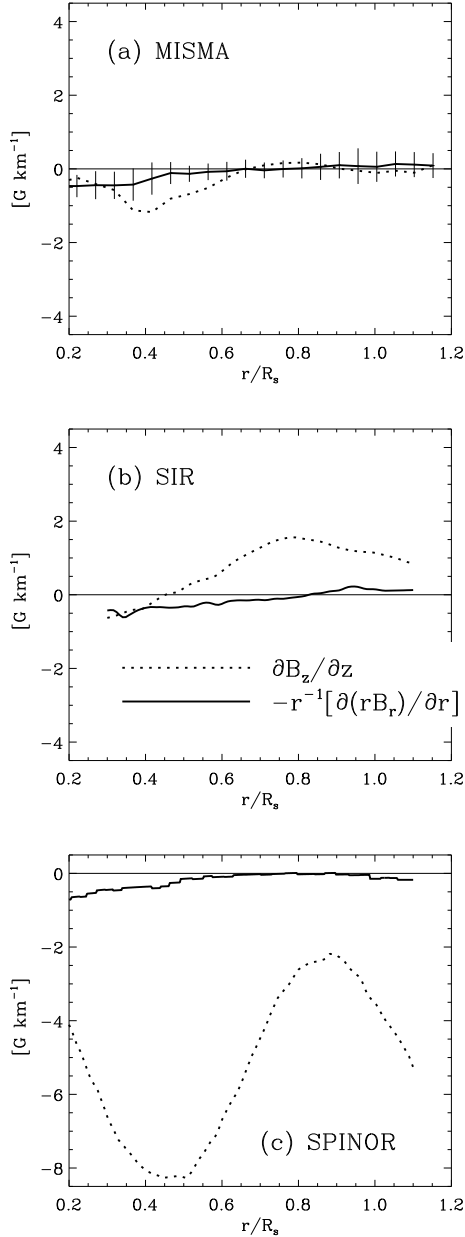}
\caption{(a) Horizontal and vertical gradients of the 
horizontal and vertical components of the retrieved 
magnetic field vector. The solid and dotted
lines represent the horizontal and 
the vertical gradients, as indicated
by the inset in (b). The error bars show
the standard deviation of the
gradients among all points
at a radial distance
from the sunspot center ($r/R_S$, with
$R_S$ the sunspot radius).
For the divergence of the magnetic field to be zero, the two
curves should be equal. The top panel shows the 
curves resulting from our inversion. For comparison,
the  panel (b) contains
the inversion by \citet{wes01a}, where 
gradients occur only along the \los .
Panel (c) shows the inversion by \citet{mat03},
which have been carried out under the same assumption.
Except for a shift, the ordinates 
of the three plots are identical.
}
\label{misma_sir1}
\end{figure}

The inversions do not provide details on the micro-structure
that give rise to the asymmetries. They are
optically-thin and the spectra are only sensitive to mean
quantities \citep{san96}. However, the condition 
$\nabla {\bf B}=0$ is also satisfied by the volume-averaged
magnetic field,
since integration and derivation are linear operators
and commute \citep{san98a}. Specifically,
\begin{equation}
\nabla <{\bf B}>=0,
	\label{div1}
\end{equation}
where the angle brackets $<~>$ are employed as in previous
sections to represent  volume average. These volume
averages are precisely the quantities  provided by the 
MISMA IC \citep{san97b}. In order to carry out 
tests, we employ equation~(\ref{div1}) integrated azimuthally.
Using cylindrical coordinates,  
equation~(\ref{div1}) becomes
\begin{equation}
	{{\partial}\over{\partial z}}\widehat{<B_z>}+
	{1\over r}
	{{\partial}\over{\partial r}}\big(r\widehat{<B_r>}\big)=0,
	\label{div2}
\end{equation}
with $B_z$ and $B_r$ the vertical and radial components
of the magnetic field,
and with the hat-symbol meaning azimuthal average, 
\begin{equation}
	\widehat{<B_z>}={{1}\over{2\pi}}\int_0^{2\pi}
		<B_z>d\chi.
   \label{div3}
\end{equation}
Figure~\ref{misma_sir1}a shows the two terms of equation (\ref{div2}).
We compute the gradients at equal continuum optical depths
($\log\tau_c=-1$). Ideally, one should
estimate horizontal gradients using magnetic fields
at a fixed geometrical
height. However, we do it at constant optical depth
keeping in mind that $\nabla<{\bf B}>=0$ must hold locally and that 
the pixel-to-pixel variation
of the Wilson depression is negligible (see Appendix \ref{appa}).
Our approach has the advantage of avoiding the
noise introduced when setting a common
scale of heights.
The sign of the vertical gradient
has been switched in Fig.~\ref{misma_sir1}
so that 
the magnetic flux is conserved when the two 
curves are identical. The two terms are indeed 
equal within a fraction of G~km$^{-1}$. 
For the sake of comparison, Fig.~\ref{misma_sir1}b
also shows the same two terms corresponding to an inversion 
where the asymmetries are 
obtained by monotonic variation of the magnetic 
field inclination along the \los . 
We derive the horizontal and vertical gradients
from Fig. 9 of \citet{wes01a}. The vertical
gradients are computed using the magnetic field 
strengths and inclinations
at two extreme optical depths ($\log\tau_c=0$
and $\log\tau_c=-2.8$) and assuming $d\tau_c/dz=-140$~km.
The horizontal gradients are computed from
the field strength and
the inclination at $\log\tau_c=-1.5$.
Figure~\ref{misma_sir1}c shows another inversion that
reproduces the Stokes asymmetries of \linc\ resorting
to monotonic variations of magnetic
field inclination along the \los .
We use Fig.~15 of \citet{mat03}, and
the above $d\tau_c/dz$. 
The MISMA inversion fulfills the condition
of magnetic flux conservation in the penumbra
 better than the 
others.
For the penumbral points, the mean absolute
values of the residuals are,
\begin{equation}
|\nabla<{\bf B}>|\sim\cases{{\rm 0.2~G~km^{-1}}& this work,\cr
				{\rm 1.2\thinspace G\thinspace km^{-1}}&
					SIR,\cr
				{\rm 4.3~G~km^{-1}}&SPINOR,}
\end{equation}
where SIR and SPINOR refer to the name of the IC
employed by \citet{wes01a} and \citet{mat03},
respectively.
%
%

The reason why MISMA inversions conform with
the magnetic flux conservation is simple. The 
gradients of magnetic field inclination
required to reproduce the Stokes~$V$ 
asymmetries are provided by the micro-structure,
rather than from systematic
variations of $<B_z>$.
There is no way two separate small and large spatial scales 
when only smooth gradients along the \los\ are permitted.
Exactly for this reason,
the same self-consistency found for
MISMAs is expected in other inversions
that allow for unresolved structure, even though
it may not necessarily be optically-thin
\citep[e.g., those with a the single fluxtube
embedded in
more vertical background;][]{bel04,bor04}

\section{Mass conservation}\label{mass}

\begin{figure}
\includegraphics[angle=0,scale=.7]{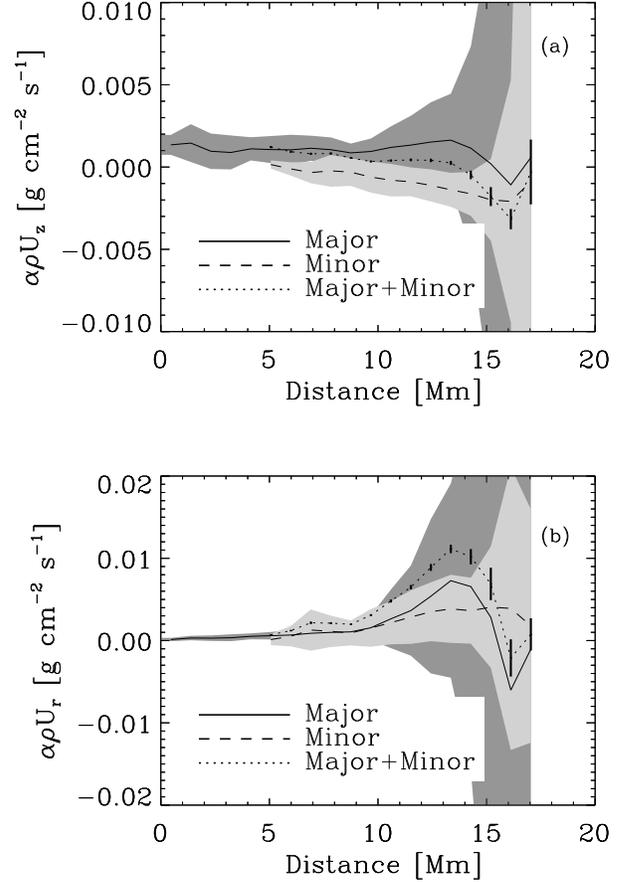}
\caption{Mass flows in the vertical direction (a; positive
upward) and  in the horizontal direction (b; positive
outward). Despite they have very different
velocities, major and minor components transport similar
amounts of mass. Note that there is no net vertical
mass flow, except at the penumbral border (a; the dotted line).
The shaded areas represent the standard deviation 
of the minor and major mass flows.
The vertical bars represent the formal error bars
of the Minor+Major mean curves (the dotted lines).
}
\label{evershed5}
\end{figure}
%
%
We have all the physical variables
required to test the model sunspot for
consistency with mass conservation.
The issue of mass conservation,
in particular at the penumbral boundary, has been
traditionally regarded as one of the difficulties-troubles-mysteries
of the Evershed  effect \citep[e.g.][]{tho94}.
Figure~\ref{evershed5} shows the mass per unit time  and
surface traveling upward 
($\alpha\rho U_z$; Fig.~\ref{evershed5}a) and
outward ($\alpha\rho U_r$; Fig.~\ref{evershed5}b). 
It includes the two MISMA components separately,
as well as net mass flux carried together.
First, note that the large velocities of the minor component 
actually transport little mass, i.e.,
similar to the mass transported by the
very modest velocities of the major component. This 
moderate mass flow results from the very low density of the minor component
(Fig.~\ref{density}) and, to a less extent, to
its reduced OCF (Fig.~\ref{ocf1}).
The mean mass transported upward by the major component
is roughly balanced by the mean mass transported downward by the minor
component. 
Figure~\ref{evershed5}b shows the horizontal mass flow.
It is positive for both the minor and the major components,
meaning that it is directed outward. This
horizontal flow decreases to zero at the penumbral boundary  
(see the dotted line in Fig.~\ref{evershed5}b
at a distance of  16~Mm). Is the mass disappearing?
The horizontal mass flux seems to turn into
downflows at the penumbral boundary.
Note how the vertical mass flow becomes negative 
at the external edge (the dotted line in
Fig.~\ref{evershed5}a at 16~Mm).
 More specifically,
the upflows associated with the major component
turns to zero so that only the minor component
downflows remain (the dashed line in Fig.~\ref{evershed5}a).
According to this picture, at the penumbral boundary
the major components cease to carry mass. We 
just see a return mass flow  similar to that
existing in the rest of the penumbra.
Formal errors for the mean net mass flows are 
included in Fig.~\ref{evershed5}. They are
computed following the recipe
in \S~\ref{mfs} and \S~\ref{evershed}.
As we point out in \S~\ref{sect_wilson},
the Wilson depression changes abruptly at the
penumbral border. This caveat must
be considered when pursuing 
the Evershed flow outside the
sunspot.

\section{The role of the minor component
	in the radiative transfer}\label{role}
\begin{figure}
\epsscale{1.0}
\plotone{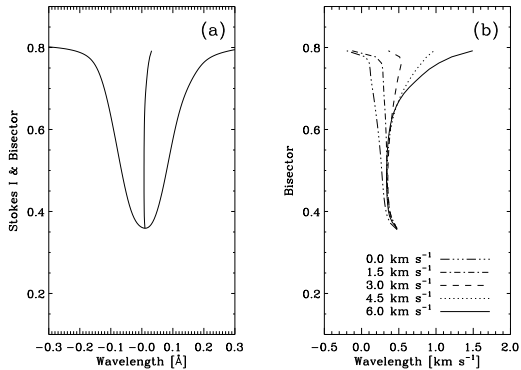}
\caption{(a) Stokes $I$ profile and its bisector for a typical
	point in the limb-side penumbra.
	(b) Bisectors corresponding to the model MISMA in (a) 
	when the velocities of the minor component
	are modified according to the legend in the figure. Note how
	the line core is insensitive to the minor component
	speed. However, the speed of the minor component
	produces significant modifications of the bisector
	at the line wings.}
\label{rt_physics}
\end{figure}

The minor component is able to modify the spectral line shapes,
despite the fact that it carries a minute fraction
of the  major component mass (about 5\%; \S~\ref{evershed}). 
This apparent contradiction admits a reasonable explanation. 
Whenever the minor component contends with the major
component to modify a line shape, it loses.
This is the reason why the minor 
component cannot modify the cores of the lines, whose
absorption is provided by the major component. 
However, due to the Doppler shift produced by
the large velocities of the 
minor component, it  often absorbs in the 
wings of the major component absorption, where
the contribution of the major component is not
so overwhelmingly dominant. Thus the minor component
can change the line wings, producing 
line asymmetries from which we infer its
presence. Figure \ref{rt_physics} tries to
illustrate this simple explanation showing how the
Stokes $I$  bisector is modified by the 
velocity of the minor component. 
The figure shows a \lina~Stokes~$I$ synthetic
profile corresponding to a point in
the limb-side penumbra. It also
includes the bisector, i.e., a line joining the mid points
of the profile at each intensity level.
The shift of the bisector is often used to measure 
velocities (e.g., \S~\ref{evershed}). Figure \ref{rt_physics}b shows 
various synthetic bisectors obtained with the same model
MISMA when the minor component velocity is varied from zero 
to the value  retrieved from the inversion\footnote{Of the 
order of 6~km~s$^{-1}$ in the layers where the line
is formed.}.
Note how the bisector position at the line core is basically
insensitive to the minor component velocity. However,
in the line wings the bisector traces   
the minor component velocity with a measurable amplitude
(2 km~s$^{-1}$).
%
%
%
%
\section{Conclusions}\label{conclusions}

Magnetic atmospheres having optically-thin micro-structure
account for the line polarization observed in plage, 
network and internetwork regions  (see the
references in \S~\ref{introduction}).
This kind of atmosphere is called  MIcro-Structured
Magnetic Atmosphere or MISMA. Our paper
shows how MISMAs also account for the 
Stokes profiles of sunspots,
including the cross-over effect 
(\S~\ref{quality}).

We have carried out a systematic 
fit of 10130 Stokes
$I$, $Q$, $U$ and $V$ profiles of \linb\ and \lina .
They correspond to the polarization emerging from
a field-of-view
containing a round sunspot  (Fig.~\ref{cont_image}).
The fits (also called inversions)
provide a model MISMA per pixel.
Each model MISMA has two magnetic
components with different magnetic fields and
mass flows interleaved along the 
line-of-sight\footnote{The fact that this MISMA scenario 
is so close to the scenarios employed
for plage, network and inter-network regions 
reflects the degeneracy of the radiative 
transfer for MISMAs \citep{san96}. The use of only two components
is an idealization that we imposed, and it actually
shows the need for more than one component to
account for the observed line shapes.} (\S~\ref{scenario}). 
Due to the nature of the radiative transfer
through
a system of optically-thin micro-structures,
such simplified model MISMAs only provide
the mean properties of the solar atmosphere. 
In particular, we are not sensitive
to the size of the individual micro-structures forming
each component (but the micro-structures
are optically-thin and so thinner than say 100~km).
By assembling 
the results of all 10130 inversions, we set up a
model sunspot with micro-structure. 
Its mean magnetic field and
mass flow is in good agreement with the standard
sunspot magnetic field
topology. The vertical magnetic fields of the 
umbra fan out and render almost horizontal fields
at the external penumbral border. Quasi horizontal
mass flows of a few km~s$^{-1}$ exist throughput
the penumbra.

This regular large scale behavior coexists
with an unexpected micro-structure. We need to
distinguish between the minor and  the
major magnetic components in  each resolution
element. The
major component is the one having most of the
mass. This terminology allows us
to describe the micro-structure of the magnetic
fields and the mass flows as follows;
\begin{enumerate}
\item The magnetic fields of the 
	minor and the major components have very different inclinations,
	with the major component being more vertical 
	than the minor component. Actually, the field lines
	of the minor component bend over to return to the 
	sub-photosphere. This return of magnetic field
	lines occurs over
	the entire penumbra, rather than at the outer penumbra
	(\S~\ref{sect_inc}; Fig.~\ref{inc_az}a).
\item By construction, the magnetic field lines and mass
	flows are always parallel. Consequently, the minor
	component carries downflows.
	According to our inversions, both upflows and
	downflows are present everywhere on the penumbra.
	The velocities along field lines are 
	small for the major component (several hundred
	m~s$^{-1}$), and large for the minor
	component (10~km~s$^{-1}$; see Fig.~\ref{evershed3a}). 
\item At equal optical depths, the major component has 
	mass densities similar to
	the unmagnetized photosphere.
	The densities of the minor component are
	 five times smaller
	(\S~\ref{sdensity}).
\item The minor component occupies a small fraction 
	of the atmosphere (about 30\% of the penumbra, 
	to become absent in the umbra; \S~\ref{ocf_sect}
	and Fig.~\ref{ocf1}).
\item	Despite the large difference of velocity, 
	the flows of the major and the minor
	components transport very similar
	masses (\S~\ref{mass}). In particular, there
	is almost no vertical transport of mass
	since the mass involved in upflows and downflows tends to be balanced.
\item The minor component has  magnetic field
	strengths  10\% larger than the
	major component. If connected by field lines,
	the resulting difference of magnetic pressure can lead to 
	a siphon flow from the major component to
	the minor component whose magnitude is similar to the
	observed Evershed flow 
	(\S~\ref{mfs}).
\item A sizeable fraction of the unsigned magnetic
	flux emerging in the penumbra returns to the 
	photosphere in~situ. In the external penumbral
	rim, the vertical magnetic flux is only
	a small fraction of the unsigned flux 
	(only 20\%).
	This result produces a considerable
	difference between the signed and unsigned 
	magnetic fluxes of the sunspot (\S~\ref{mfs}).
\item The model sunspot has a Wilson depression
	of $-0.25$~Mm 
	at the solar disk center,
	and it significantly increases toward
	the limb (\S~\ref{sect_wilson}).
%
%
\end{enumerate}

Counter-intuitive as it may sound,  the presence of
field lines pointing up and down over the entire
penumbra does not obviously conflict with 
existing observations.
We have carried out several 
tests to check the internal consistency 
of the retrieved model sunspot. 
\begin{itemize}
\item The MISMA inversions account for the asymmetries
	of the Stokes profiles
	with a magnetic field vector
	{\bf B} which does not violate the $\nabla {\bf B}=0$ condition
	(\S~\ref{mfc_sect}). 
\item Although, the model sunspot has a Wilson depression
	smaller than the typical values found
	in the literature,
	it reproduces the observables from which large Wilson 
	depressions are inferred
	(\S~\ref{sect_wilson}).  
\item The model sunspot reproduces the Net Circular Polarization
	(NCP) of the observed lines,
	including their rules. In addition, it also
	reproduces the abnormal behavior of the
	NCP of \linc , discovered and explained by 
	Schlichenmaier and co-workers using
	magnetic structures larger than the ones implicit
	in the MISMA assumption (\S~\ref{ncp_sect}).
\item	When the observed Stokes profiles are interpreted
	assuming a single magnetic component per resolution
	element, one finds upflows in the inner penumbra and
	downflows in the outer penumbra 
	(\S~\ref{evershed}; Fig.~\ref{evershed4}). These 
	upflows and downflows appear in the 
	recent penumbral literature. According to
	our interpretation, they are artificially
	produced when the unresolved structure of the flows 
	is neglected (\S~\ref{evershed}). 
	It would be similar to the 
	well known convective blueshift of the non-magnetic 
	granulation, which does not correspond to a true
	systematic motion.
\item Despite of the partial cancellation of vertical 
	magnetic flux in the penumbra, some
	net magnetic flux remains.
	The existence of such residual flux
	avoids an obvious disagreement of the model
	sunspot with existing 
	observations. Chromospheric and coronal
	images often show fieldlines 
	connecting the penumbra with magnetic
	structures far away from the sunspot
	\citep[see, e.g.,][]{wei04}.
\item	The minor component contributes to the
	emerging spectrum
	despite the fact that they
	contain a minute fraction of the total penumbral
	mass (5\% or
	less). Such a disproportionate effect results
	from the large Doppler shifts associated with the minor
	components, which absorb in
	the wings of the major component 
	absorption profiles (\S~\ref{role}).
\end{itemize}

The interpretation of all these results in the 
context of a coherent sunspot magnetic structure is  
deferred for a 
separate  paper \citep{san04c}.
We prefer to disconnect the fact that
MISMAs account for the Stokes asymmetries
observed in sunspots (this work), from the
more speculative physical
interpretation of the retrieved 
micro-structure. 
Even if the physical interpretation 
is eventually shown to be wrong, 
MISMAs would still account for the asymmetries. 
Only one general remark concerning the 
global magnetic structure
will
be made. The mass flows and 
magnetic fluxes of the minor and major 
components are surprisingly similar. 
This
suggests a physical connection between them, so
that field lines and flows emerging as the major component 
return to the photosphere as the minor component
of the same or a neighbor resolution element. This possibility
is not a direct result of the MISMA inversion, and it is put forward
as a mere conjecture.

	The picture that emerges from our sunspot inversions
	is in qualitative agreement with the original
	MISMA scenario for penumbrae put forward by
	\citet[][ \S~3.1]{san96} 

A final caveat is in order. 
The fact that our model sunspot
reproduces the observed Stokes asymmetries does 
not guarantee the accuracy of the model.
However, that fact that a model sunspot reproduces the asymmetries
should not be undervalued. 
If the model sunspot that accounts
for real sunspots is eventually discovered, it will be found 
among those in this rare
category of models reproducing the Stokes asymmetries.	


\acknowledgements
Special thanks are due to B. Lites and J. Thomas,
who supplied
the data analyzed in this work and provided
insightful comments on an earlier version of the 
manuscript.
The work has partly been funded by the Spanish Ministry of Science
and Technology, 
project AYA2001-1649, as well as by
the EC contract HPRN-CT-2002-00313.
%
%
%

%
%
\appendix
\section{Error of the absolute scale of heights}\label{appa}

An absolute scale of heights is derived assuming 
no variation of the total pressure 
between adjacent pixels (\S~\ref{scale_of_height}).
The purpose of this appendix is to estimate the error
introduced by such assumption.

The true total pressure $P$ depends not only 
on the height in the atmosphere, $z$, but also on  the
horizontal coordinates. For the sake of simplicity,
we consider an axi-symmetric sunspot so that the
horizontal coordinates are reduced to
the distance to the sunspot center
$r$. 
Let us use $\widetilde P(r,z)$ to denote
the distribution of pressures inferred assuming 
that there is no spatial variation of $P(r,z)$.
The estimated pressures are identical to the true
pressures except for a shift of the vertical scale,
\begin{equation}
\widetilde P(r,z)=P\big(r,z-z_0(r)\big).
\end{equation}
This shift, $z_0(r)$, is the error of the vertical
scale that we want to estimate.
The $\widetilde P(r,z)$ has been constructed
so that its variation with the horizontal position is
minimum, i.e.,
\begin{equation}
{{d\widetilde P(r,z)}\over{d r}}=
{{\partial P\big(r,z-z_0(r)\big)}\over{\partial r}}
-{{\partial P\big(r,z-z_0(r)\big)}\over{\partial z}}
{{dz_0(r)}\over{dr}}
\simeq 0.
\end{equation}
This expression provides an ordinary differential equation for
the error,
\begin{equation}
{{dz_0(r)}\over{dr}}\simeq 
{
{{\partial \ln P\big(r,z-z_0(r)\big)}/{\partial r}}
\over
{{\partial \ln P\big(r,z-z_0(r)\big)}/{\partial z}}
},
\end{equation}
which can be integrated to evaluate $z_0$.
The right-hand-side of the previous equation
can be estimated from
the magnetostatic equilibrium equation. It couples
the spatial gradients of gas pressure $P_g$, radial
magnetic field $B_r$, and 
vertical magnetic field $B_z$ \citep[e.g.,][ \S 8.4]{pri82},
\begin{equation}
{{\partial P_g}\over{\partial r}}
={{B_z}\over{4\pi}}\Big({{\partial B_r}\over{\partial z}}
-{{\partial B_z}\over{\partial r}}\Big).
	\label{appa1}
\end{equation}
Using the conservation of magnetic flux in 
an axi-symmetric sunspot with radial magnetic fields,
\begin{equation}
	{{\partial B_z}\over{\partial z}}+
{1\over r}{{\partial(r B_r)}\over{\partial r}}=0,
\end{equation}
and the definition of total pressure,
\begin{equation}
P=P_g+{{B_r^2}\over{8\pi}}+{{B_z^2}\over{8\pi}},
\end{equation}
it is possible to transform equation~(\ref{appa1})
into 
\begin{equation}
{{\partial P}\over{\partial r}}=-{{B_r^2}\over{4\pi r}}
	+{{B^2}\over{4\pi}}{{\partial \iota}\over{\partial z}},
	\label{master}
\end{equation}
where $\iota$ stands for the magnetic field inclination with
respect to the vertical direction (Fig.~\ref{schematic1}).
(Note that we drop from the expressions
the dependence of the physical
variables on height and radial position. We do it for the
sake of simplicity, without compromising clarity.)
The second term of the right-hand-side of equation (\ref{master})
vanishes for the
kind of atmosphere that we deal with (whose mean
vertical  gradient of inclination is zero), therefore,
\begin{equation}
{{\partial\ln P}\over{\partial r}}\simeq
-{2 \over{1+\beta}}{{\sin^2\iota}\over r},
\end{equation}
where we use  $\beta$ for the ratio of gas pressure
to magnetic pressure,
$\beta=8\pi P_g/B^2$. Using this equation plus the 
definition of the total pressure scale height $H_P$,
\begin{equation}
{{\partial\ln P}\over{\partial z}}\equiv -H_P^{-1},
\end{equation}
one can find a very compact expression for the 
horizontal gradient of the  error,
\begin{equation}
{{dz_0(r)}\over{dr}}\simeq 
{2 H_P \over{1+\beta}}{{\sin^2\iota}\over r}.
	\label{this_zero}
\end{equation}
This differential equation can be easily used 
to estimate both,
the error between adjacent pixels separated by
$\delta r$,
\begin{equation}
\delta z_0(r)\simeq\delta r~{{dz_0(r)}\over{dr}},
\label{this_one}
\end{equation}
and the total error from the sunspot center 
to a given radial distance $r$,
\begin{equation}
\Delta z_0(r)=\int_0^r{{dz_0(r')}\over{dr}} dr'.
\label{this_two}
\end{equation}

\begin{figure}
\epsscale{0.5}
\plotone{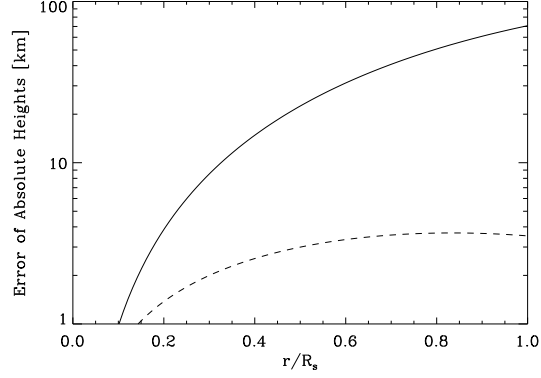}
\caption{Error of the absolute vertical scale 
due to the assumption  that the total pressure 
does not vary between neighbor pixels. They are
represented for each radial position along the
sunspot ($r/R_S$, with $R_S$ the sunspot radius).
The solid line is the error relative to
the sunspot center, whereas the dashed line correspond
to the error between two adjacent pixels.
Heights are given in km.
}
\label{appa_fig1}
\end{figure}

Figure \ref{appa_fig1} shows both $\delta z_0$ and
$\Delta z_0$ computed according to equations~(\ref{this_zero}), 
(\ref{this_one}), and (\ref{this_two}),
under the assumption 
$\beta= 1$, $H_P= 100~$km, $\delta r\simeq 0.75\arcsec$, and
\begin{equation}
\iota=80^\circ {r\over{R_S}},
\label{this_inc}
\end{equation}
with $R_S$ the sunspot radius. 
Equation (\ref{this_inc})
roughly describes the mean inclination 
of a sunspot magnetic field (e.g., \citealt{sol00}, or Fig.~\ref{inc_az}).
The rest of parameters are typical of our penumbral
inversions. Note that 
$\sin^2\iota/r\propto r$ in the umbra,  so that integrand of 
equation (\ref{this_two}) is only significant in the penumbra.
According to Fig.~\ref{appa_fig1},
the pixel-to-pixel error is negligible (smaller
than 4 km). The variation across the full spot is larger,
but it does not reach 100~km.

\end{document}